\documentclass[acmsmall]{acmart}
\usepackage[T1]{fontenc}
\usepackage[utf8]{inputenc}
\usepackage[english]{babel}
\usepackage{microtype}
\usepackage{pifont}

\usepackage{amsmath}
\DeclareMathOperator{\avg}{avg} 

\usepackage{algorithm}
\usepackage{algorithmicx}
\usepackage[noend]{algpseudocode}

\AtBeginDocument{%
\setlength{\belowdisplayskip}{2pt}
\setlength{\belowdisplayshortskip}{2pt}
\setlength{\abovedisplayskip}{3pt}
\setlength{\abovedisplayshortskip}{3pt}
}

\usepackage{booktabs,multirow,array,ragged2e}
\usepackage{graphicx}
\usepackage[caption=false,font=footnotesize,subrefformat=parens,labelformat=parens]{subfig}
\usepackage{capt-of}
\usepackage{rotating}
\usepackage{adjustbox}
\usepackage{xcolor}
\usepackage{pdflscape}

\usepackage[authormarkup=superscript,deletedmarkup=sout,addedmarkup=em]{changes}
\usepackage{soul}
\soulregister\cite7
\soulregister\ref7
\soulregister\pageref7

\colorlet{soulcyan}{cyan!30}
\colorlet{soulgreen}{green!30}

\definechangesauthor[name={Claudio}, color=green!75!black]{CF}
\definechangesauthor[name={Salvo}, color=magenta!75!black]{SD}

\usepackage{siunitx}
\usepackage{xspace}
\usepackage{enumitem}

\newcommand{\eg}{\textit{e.g.},\xspace}
\newcommand{\ie}{\textit{i.e.},\xspace}

\newcommand{\simpletitle}[1]{\noindent\textbf{#1}.\xspace}

\newcommand{\oran}{O-RAN\xspace}

\newcommand\txbrate{\relax\ifmmode\mathtt{tx\_bit\-rate}\else\texttt{tx\_bit\-rate}\fi\xspace}
\newcommand\txpkts{\relax\ifmmode\mathtt{tx\_packets}\else\texttt{tx\_packets}\fi\xspace}
\newcommand\dlbuff{\relax\ifmmode\mathtt{DWL\_buffer\_size}\else\texttt{DWL\_buffer\_size}\fi\xspace}

\usepackage[printonlyused]{acronym}

\acrodef{ai}[AI]{Artificial Intelligence}
\acrodef{xai}[XAI]{EXplainable Artificial Intelligence}
\acrodef{ml}[ML]{Machine Learning}
\acrodef{dl}[DL]{Deep Learning}
\acrodef{aml}[AML]{Adversarial Machine Learning}
\acrodef{nn}[NN]{Neural Networks}
\acrodef{dnn}[DNN]{Deep Neural Networks}
\acrodef{lstm}[LSTM]{Long-Short Term Memory}
\acrodef{gnn}[GNN]{Graph Neural Networks}
\acrodef{gan}[GAN]{Generative Adversarial Networks}
\acrodef{mae}[MAE]{Mean Absolute Error}
\acrodef{mse}[MSE]{Mean Square Error}
\acrodef{shap}[SHAP]{SHapely Additive exPlanations}
\acrodef{lime}[LIME]{Local Interpretable Model-agnostic Explanations}
\acrodef{xgb}[XGBoost]{Extreme Gradient Boosting}
\acrodef{lrp}[LRP]{LayeR-wise backPropagation}
\acrodef{fgsm}[FGSM]{Fast Gradient Sign Method}
\acrodef{bim}[BIM]{Basic Iterative Method}
\acrodef{dt}[DT]{Decision Tree}
\acrodefplural{DT}[DTs]{Decision Trees}

\acrodef{xrl}[XRL]{EXplainable Reinforcement Learning}
\acrodef{rl}[RL]{Reinforcement Learning}
\acrodef{drl}[DRL]{Deep Reinforcement Learning}
\acrodef{crl}[CRL]{Casual Reinforcement Learning}
\acrodef{dqn}[DQN]{Deep Q-Network}
\acrodef{ppo}[PPO]{Proximal Policy Optimization}
\acrodef{a3c}[A3C]{Asynchronous Advantage Actor-Critic}

\acrodef{ran}[RAN]{Radio Access Network}

\acrodef{mno}[MNO]{Mobile Network Operator}
\acrodef{bs}[BS]{Base Station}
\acrodefplural{BS}[BSs]{Base Stations}
\acrodef{gnb}[gNB]{next Generation Node B}
\acrodefplural{gNB}[gNBs]{next Generation Node Bs}
\acrodef{enb}[eNB]{evolved Node B}
\acrodefplural{eNB}[eNBs]{evolved Node Bs}
\acrodef{ue}[UE]{User Equipment}

\acrodef{prb}[PRB]{Physical Resource Block}
\acrodefplural{PRB}[PRBs]{Physical Resource Blocks}

\acrodef{embb}[eMBB]{enhanced Mobile BroadBand}
\acrodef{mmtc}[mMTC]{massive Machine-type Communications}
\acrodef{urllc}[URLLC]{Ultra-Reliable and Low Latency Communications}
\acrodef{mcs}[MCS]{Modulation and Coding Scheme}

\acrodef{qos}[QoS]{Quality of Service}
\acrodef{kpi}[KPI]{Key Performance Indicator}
\acrodefplural{kpi}[KPIs]{Key Performance Indicators}
\acrodef{sla}[SLA]{Service Level Agreement}
\acrodef{isp}[ISP]{Internet Service Providers}
\acrodef{uav}[UAV]{Unmanned Aerial Vehicle}
\acrodef{mdp}[MDP]{Markov Decision Process}

\acrodef{ric}[RIC]{RAN Intelligent Controller}
\acrodef{smo}[SMO]{Service Management and Orchestrator}

\acrodef{srn}[SRN]{Standard Radio Node}
\acrodef{rmr}[RMR]{RIC Message Router}

\acrodef{ht}[HT]{High-Throughput}
\acrodef{ll}[LL]{Low-Latency}

\acrodef{dwl}[DWL]{downlink}

\acrodef{abr}[ABR]{Adaptive Bit Rate}

\usepackage{xurl}

\usepackage{dblfloatfix}

\usepackage{orcidlink}

\newcommand{\toolname}{\textit{EXPLORA}\xspace}

\usepackage{tikzpagenodes,etoolbox}
\usepackage[contents={}]{background}
\AddEverypageHook{%
	\ifnumequal{\thepage}{1}{%
		\tikz[remember picture,overlay]{%
			\node[draw,
			minimum width=\textwidth,
			text width=0.99\textwidth,
			font=\footnotesize,
			] 
			at ([yshift=6pt]current page header area) 
			{%
			This is the author's accepted version of the article. The final version published by ACM is C. Fiandrino, L. Bonati, S. D'Oro, M. Polese, T. Melodia and Joerg Widmer, ``\textit{EXPLORA}: AI/ML \textit{EXPL}ain\-ability for the \textit{O}pen~\textit{RA}N,'' \textit{ACM International Conference on emerging Networking EXperiments and Technologies (CoNEXT)}, Paris, France, 2023, pp. TBD, doi: TBD.
			};
		}%
	}{}
}

\begin{document}

\title{\textit{EXPLORA}: AI/ML \textit{EXPL}ain\-ability for the \textit{O}pen~\textit{RA}N}

\author[C. Fiandrino]{Claudio Fiandrino\,\orcidlink{0000-0002-4323-4355}}
\email{claudio.fiandrino@imdea.org}
\orcid{0000-0002-4323-4355}
\affiliation{%
\institution{IMDEA Networks Institute}
\streetaddress{}
\city{Madrid}
\country{Spain}
}

\author[L. Bonati]{Leonardo Bonati\,\orcidlink{0000-0002-1511-1833}}
\email{l.bonati@northeastern.edu}
\orcid{0000-0002-1511-1833}
\affiliation{%
\institution{Institute for the Wireless Internet of Things, Northeastern University}
\streetaddress{}
\city{Boston}
\country{USA}
}

\author[S. D'Oro]{Salvatore D'Oro\,\orcidlink{0000-0002-7690-0449}}
\email{s.doro@northeastern.edu}
\orcid{0000-0002-7690-0449}
\affiliation{%
\institution{Institute for the Wireless Internet of Things, Northeastern University}
\streetaddress{}
\city{Boston}
\country{USA}
}

\author[M. Polese]{Michele Polese\,\orcidlink{0000-0002-9740-134X}}
\email{m.polese@northeastern.edu}
\orcid{0000-0002-9740-134X}
\affiliation{%
\institution{Institute for the Wireless Internet of Things, Northeastern University}
\streetaddress{}
\city{Boston}
\country{USA}
}

\author[T. Melodia]{Tommaso Melodia\,\orcidlink{0000-0002-2719-1789}}
\email{melodia@northeastern.edu}
\orcid{0000-0002-2719-1789}
\affiliation{%
\institution{Institute for the Wireless Internet of Things, Northeastern University}
\streetaddress{}
\city{Boston}
\country{USA}
}

\author[J. Widmer]{Joerg Widmer\,\orcidlink{0000-0001-6667-8779}}
\email{joerg.widmer@imdea.org}
\orcid{0000-0001-6667-8779}
\affiliation{%
\institution{IMDEA Networks Institute}
\streetaddress{}
\city{Madrid}
\country{Spain}
}

\begin{abstract}
The Open \ac{ran} paradigm is transforming cellular networks into a system of disaggregated, virtualized, and software-based components. These self-optimize the network through programmable, closed-loop control, leveraging \ac{ai} and \ac{ml} routines. In this context, \ac{drl} has shown great potential in addressing complex resource allocation problems. However, \ac{drl}-based solutions are inherently hard to explain, which hinders their deployment and use in practice. In this paper, we propose \toolname, a framework that provides explainability of \ac{drl}-based control solutions for the Open \ac{ran} ecosystem. \toolname synthesizes network-oriented explanations based on an attributed graph that produces a link between the actions taken by a \ac{drl} agent (\ie the nodes of the graph) and the input state space (\ie the attributes of each node). This novel approach allows \toolname to explain models by providing information on the wireless context in which the \ac{drl} agent operates. \toolname is also designed to be lightweight for real-time operation.
We prototype \toolname and test it experimentally on an O-RAN-compliant near-real-time RIC deployed on the Colosseum wireless network emulator.
We evaluate \toolname for agents trained for different purposes and showcase how it generates clear network-oriented explanations. We also show how explanations can be used to perform informative and targeted intent-based action steering and achieve median transmission bitrate improvements of $4$\% and tail improvements of $10$\%.
\end{abstract}


\maketitle

\acresetall

\section{Introduction}
\label{sec:intro}

To support this rapidly changing and complex environment foreseen in the sixth generation (6G), the industry is now transitioning toward \ac{ran} architectures based upon softwarization, virtualization, and network programmability paradigms, such as the Open \ac{ran}~\cite{comst-surveoran-wines}. Specifying how to practically realize an Open RAN architecture is one of the goals of the O-RAN Alliance. O-RAN leverages the above principles to provide an alternative to existing inflexible, monolithic equipment with systems based on disaggregated, virtualized, and software-based components that interact via open and standardized interfaces. O-RAN also introduces two \acp{ric} that act as abstraction layers to monitor, control and manage \ac{ran} components at different timescales, namely at near-real-time (or near-RT) and non-real-time (non-RT) timescales. The near-RT and non-RT \acp{ric} host custom applications, respectively xApps and rApps, that run \ac{ai}-based closed-loop control routines to optimize the \ac{ran} operations and adapt them to current traffic demand and network conditions~\cite{comst-surveoran-wines}.

Bootstrapped by such initiatives, the application of \ac{ai} to the Open \ac{ran} has become an emerging area of interest~\cite{comst-surveoran-wines}, with contributions that encompass spectrum management~\cite{milcom21-ss,infocom22-luca-wines}, mobility management~\cite{mdpisensors-trafficsteeringoran}, and resource allocation~\cite{tmc-coloran-wines,wintech-nextran,csm21-andres}, as well as custom control loops to jointly optimize location and transmission directionality of \acp{uav}~\cite{tmc-lorenzo-wines}. Among the existing data-driven techniques, \ac{drl} appears to be particularly suitable to control Open \ac{ran} systems~\cite{li2021rlops}.
Unlike supervised learning models--- tailored to classification or regression tasks---\ac{rl} and \ac{drl} focus on decision-making processes where decisions are made through a trial-and-error process to maximize a certain utility function (\eg the throughput of the network). \ac{drl} is widely used for several networking problems related to resource allocation, handover and load balancing~\cite{comst-drl-networking,xu-drl-infocom18,li2023tapfinger,ho2022joint}, among others.
Because of its capability of interacting with, and adapting to, complex, highly distributed, dynamic and uncertain environments---such as those typical of cellular \acp{ran}---is a compelling candidate AI technique for Open \ac{ran}. Successful industry applications of \ac{drl} in the \ac{ran} cover relevant use cases such as traffic steering (Mavenir~\cite{mavenir-drl}), and handover management (Intel~\cite{intel-drl}), among others.

\simpletitle{Motivations and Objectives}~Since \ac{drl} agents leverage deep neural networks, the logic governing their decisions is frequently hard to understand. This is unlike, for example, \acp{dt}~\cite{lundberg2020local}, whose structure and decision-making logic is generally explicit and easy to understand, especially in relatively simple and confined practical applications such as automated \ac{bs} reconfiguration~\cite{auric}. Despite their effectiveness, the lack of explainability of \ac{drl} models makes them difficult to use in production networks because of the inherent lack of understanding of the logic behind decisions. This complicates troubleshooting and predicting decisions, and makes \ac{drl} more vulnerable to adversarial attacks~\cite{deexp23}. The issue affects \ac{ai} at large, not only \ac{drl}, and makes understanding why models take certain actions more difficult, especially with complex environments and large action spaces.

To address these problems, the research community is trying to shed light on the inner mechanisms of such models to make them more explainable and interpretable.
For instance, Puiutta et al.~\cite{puiatta-xrl-survey} coined the term \ac{xrl} and illustrate several explainability techniques that are specific to the learning paradigm. However, although the interest in promoting trust and interpretability to AI has recently gained momentum~\cite{arxiv-survey}, explainable AI in the context of mobile networks is still at its early stages and largely unexplored~\cite{comcom-xai,zili-interpret-sigcomm20}. 

\simpletitle{Our Contribution} In this paper, we try to fill this gap and make \ac{drl} for Open \ac{ran} applications more robust, resilient and---more importantly--- explainable. Specifically, we develop \toolname, a lightweight O-RAN-compliant framework designed to explain the logic of \ac{drl} agents executed within xApps or rApps performing near- and non-real-time closed-loop resource allocation and control. In contrast with the traditional approach, where model explanations only provide intuitions that reveal \textit{how inner mechanisms of a model work} (\eg neuron activation), we ensure that \toolname also provides informative explanations on the wireless network behavior to help operators in interpreting AI decisions~\cite{zili-interpret-sigcomm20}. Specifically, we advance \ac{xai} and \ac{xrl} research in several ways. 

We show that \toolname addresses complex challenges (\S~\ref{subsec:challenges}) that prevent the direct use of state-of-the-art \ac{xai} tools (\S~\ref{subsec:motivation-shap}). We base the \ac{xai} component of \toolname on attributed graphs to connect the actions taken by the agents (nodes) to the effect on the environment (attributes of the nodes) and to distill knowledge by analyzing the transitions between actions (edges - \S~\ref{sec:interpretation}). This allows \toolname to explain models by highlighting the circumstances under which a \ac{drl} agent takes specific actions. Overall, such combined knowledge (\ie input-output relations and conditions that trigger certain actions) is useful to domain experts and mobile operators willing to retain full control of the \ac{ai}-based system and, if needed, to consciously override or inhibit \ac{ai} decisions on the basis of previously identified intents to be fulfilled.

We experimentally demonstrate the explainability capabilities of \toolname on Colosseum, the world's largest O-RAN wireless network emulator~\cite{colosseum}. Specifically, we apply \toolname to xApps embedding \ac{drl} agents for control of \ac{ran} slicing and scheduling, developed via the OpenRAN Gym open-source O-RAN data collection and AI testing toolbox~\cite{openrangym} (\S~\ref{subsec:impl-packaging}). We show not only that \toolname is capable of synthesizing explanations that facilitate monitoring and troubleshooting (\S~\ref{subsec:results-explanations}), but also that it helps to improve \ac{ran} performance by using explanations to proactively identify and substitute actions that could lead to poor performance (\S~\ref{subsec:results-optimization}).

\simpletitle{Key Contributions and Findings} Our far-reaching goal is to contribute toward promoting \ac{ai} trustworthiness in Open \ac{ran}. The key contributions (marked with ``C'') and findings (``F'') of our study are summarized as follows:%
\vspace*{-1ex}
{%
\setlist[itemize]{leftmargin=1.75em}
\begin{itemize}
\item[C1.] We propose \toolname, a new framework for network-oriented explanations. \toolname provides informative post-hoc explanations and can evaluate the effectiveness of control actions taken by the \ac{drl} agents;
\item[C2.] We implement \toolname as an xApp on an O-RAN-compliant near-RT \ac{ric}; and 
\item[C3.] We release the artifacts of our study on \url{https://github.com/wineslab/explora}.
\item[F1.] We find that \toolname provides effective and concise explanations fully characterizing \ac{drl} agents behavior.
\item[F2.] We find that \toolname enables the creation of ad-hoc policies for intent-based action steering that ultimately improve users' \acp{kpi}. We observe median transmission bitrate improvements of $4$\% and tail improvements of $10$\%.
\end{itemize}
}

\section{Background}
\label{sec:bck}

In this section, we provide background knowledge on the different technologies considered in our paper: O-RAN (\S~\ref{subsec:bck-oran}), \ac{drl} (\S~\ref{subsec:bck-drl}), and finally \ac{xai} and \ac{xrl} (\S~\ref{subsec:bck-xai}).

\subsection{Background on O-RAN}
\label{subsec:bck-oran}

\begin{figure}[tbp]
\centering%
\includegraphics[width=.35\textwidth, keepaspectratio]{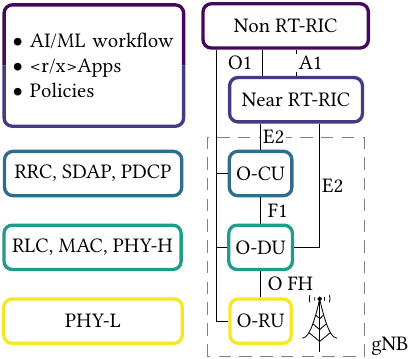}%
\vspace*{-2ex}%
\caption{The O-RAN reference architecture}%
\label{fig:oran-ref-arch}%
\vspace*{-6ex}%
\end{figure}

The \oran specifications introduce a complete architectural model for the Open \ac{ran} (see Figure~\ref{fig:oran-ref-arch}). The interactions and interoperability between multi-vendor equipment implementing disaggregated \ac{ran} \acp{gnb} (\ie central, distributed and remote units---O-CU, O-DU and O-RU) is possible via open and standardized interfaces. Management and control is provided by a set of \acfp{ric} that operate at different timescales, \ie non-RT (timespan larger than $1$~s) and near-RT (timespan between $10$~ms to $1$~s)~\cite{comst-surveoran-wines}. The non-RT RIC enforces policies controlling thousands of devices, including data collection and training phase of the \ac{ai}/\ac{ml} workflows at large and provides the near-RT RIC with policy-based guidance through the A1 interface. It is embedded in the network \ac{smo}, which performs automated monitoring and provisioning of network functions through the O1 interface. The near-RT RIC operates control loops for policy enforcement (\ie control) at a smaller scale (tens to hundreds of nodes) and governs radio resource management operations such as resource allocation~\cite{comst-surveoran-wines, csm21-andres} by interacting with the \ac{ran} nodes through the E2 interface. The RICs can host third-party applications, \ie rApps at non-RT scale and xApps at near-RT scale. These custom applications execute control logic for dynamic network optimization and are a key enabler for enforcing zero-touch network automation and self-configuration.

\subsection{Background on DRL}
\label{subsec:bck-drl}
In \ac{rl} and \ac{drl}, agents dynamically interact with an \textit{environment} to maximize a target utility function~\cite{sutton2018reinforcement}. Formally, the environment is defined as a \ac{mdp}. At each time step $t$, the agent first observes the environment to estimate its \textit{state} $s_t \in \mathcal{S}$, and then takes an \textit{action} $a_t \in \mathcal{A}$ that follows a \textit{policy} \mbox{$\pi:\mathcal{S}\rightarrow P(\mathcal{A})$}. In general, actions can be multi-modal and consist of combined decisions affecting two or more control parameters. Specifically, any action $a_t$ can be expressed as a $c$-tuple, \ie \mbox{$a_t=(a_t^1,a_t^2,\dots,a_t^c)$}, where $c$ represents the number of modes of the action. When the action $a_t$ is taken, the environment transitions from state $s_t$ to the next state $s_{t+1}$ following a transition kernel $T(s_{t+1}|s_t,a_t)$, and generates a \textit{reward} $r_t: \mathcal{S}\times \mathcal{A} \rightarrow \mathbb{R}$. Starting from an initial state $s_0$, the tuple $M=(\mathcal{S}, \mathcal{A}, T, r, s_0, \gamma)$ defines the \ac{mdp} and, in the most general setup, the objective of the agent is to learn a policy $\pi$ that maximizes the weighted reward \mbox{$R(\pi) =  \sum_{t=0}^{+\infty}{\gamma^t \, r_t}$}, where $\gamma$ is a discount factor used to fine-tune the weighted sum of rewards. Specifically, a value of $\gamma$ closer to $1$ would aim at maximizing the long-term reward, while a value closer to $0$ would prioritize the short-term reward.

\subsection{Background on XAI and XRL}
\label{subsec:bck-xai}

\simpletitle{A Primer on \ac{ai} Explainability} Promoting trustworthiness in \ac{ai} has received a lot of interest in recent years~\cite{gunning2019darpa,regulating-ai,arxiv-survey}. While \textit{interpretability} contextualizes the model decision-making in relation to its internal design, \textit{explainability} encompasses \textit{interpretability} and goes beyond by aiming at justifying how and why a model achieves a given output in a human understandable manner~\cite{broniatowski2021psychological}.

Regarding interpretability, there exist model-agnostic and model-specific \ac{xai} techniques. Both reveal which part of the input data or features have more influence on a prediction. \ac{shap}~\cite{shap-lundberg2017unified}, \ac{lime}~\cite{lime-kdd}, and Eli5~\cite{eli5} are model-agnostic and provide model interpretations by identifying input relevance via perturbation techniques. Instead, DeepLIFT~\cite{deeplift} and \ac{lrp}~\cite{montavon-lrp} provide interpretations by evaluating which activation/neurons were relevant to a prediction via backpropagation. Hence, these techniques need to be specialized for the specific \ac{ai} model.

\simpletitle{\ac{xai} and \ac{xrl}} The literature is inconsistent in the way the terms are classified:~\cite{krajna-xrl} defines \ac{xai} and \ac{xrl} as a collection of techniques applicable to supervised and \ac{rl} respectively, while~\cite{dazeley-xrl, puiatta-xrl-survey} take a different approach and define \ac{xrl} as a subset of \ac{xai}. We find the latter definition more appropriate as techniques like \ac{shap} and \ac{lrp} have been applied to \ac{drl}~\cite{zhang-shap-drl-power-tcss,xu-lrp-drl}.

\section{Motivation and Challenges}
\label{sec:bck-motiv}

In this section, we first delve into the design principles of \ac{drl} solutions for the \ac{ran}, and then illustrate why currently available \ac{xai} tools cannot be applied \textit{as-is} to such \ac{drl} models.

\simpletitle{Data-driven \ac{ran} reconfiguration} The \ac{ran} is characterized by a large number of parameters and functionalities that can be monitored and configured. In general, these can be categorized according to the timescale at which they are updated. Cell parameters like cell ID, coverage radius, antenna orientation and energy saving mechanisms are usually updated over the course of several seconds, minutes, hours or even days. In contrast, transmission power control policies, interference management, handover, and resource allocation are configured in the sub-second timescale. How to optimally determine these configurations is a well-known problem which has been tackled several times via \acp{dt} or \ac{drl}. \acp{dt} work well in the case of feature selection or rule-based configuration empowered by past historical data like the case of Auric for \acp{bs}~\cite{auric} or Configanator for content-delivery networks~\cite{naseer2022configanator}. \ac{drl} agents are effective solutions for configuring parameters at shorter timescales where dynamic reconfiguration must be achieved by adapting and responding to real-time measurements. Alternatively, they can be used after the exploitation stage for parameter configuration~\cite{ge-chroma-mobicom23}. \acp{dt} are self-interpretable vis-a-vis with \ac{drl} agents and have been used either directly for rule-based configurations~\cite{auric,naseer2022configanator}, or indirectly to deliver explanations of \ac{drl} agents that enforce simple decisions like bitrate selection in \ac{abr} context~\cite{zili-interpret-sigcomm20}.

Unfortunately, \ac{drl} solutions for O-RAN systems are, in general, more complex than those used in the above examples, and they share a common set of features that we elaborate hereafter.
\textit{First}, O-RAN networks belong to a class of systems that are very hard to model and observe. Thus, it is common practice to feed \ac{drl} agents with a low-dimensionality latent representation of the observed network state rather than with the state itself.
Indeed, the latter usually consists of large quantities of heterogeneous and real-time \ac{kpi} measurements with high variance, which might results in the so-called \textit{state space explosion}, where the number of states is so large that the agents cannot learn an effective policy or would require excessively long training time.
The use of autoencoders is a well-established \ac{ml} tool to mitigate the above issues~\cite{ac-for-drl,ac-drl-example}.
\textit{Second}, \ac{drl} agents may take hierarchical actions where controllable parameters depend on the value of other non-controllable parameters, previously observed states, or actions taken in the past (\eg a \ac{drl} agent controlling resource allocation policies subject to higher level power control~\cite{iturria2022multi}). 
\textit{Third}, \ac{drl} agents may take multi-modal actions that involve diverse control parameters. Practical examples include making joint decisions on user scheduling and antenna allocation~\cite{huang-access-complex-action}, \ac{ran} scheduling and slicing policies~\cite{tmc-coloran-wines}, or simultaneous control of computing resources and \ac{mcs}~\cite{vrain-andres-mobicom19}.

\subsection{Use Case Throughout the Paper}
\label{subsec:drl-use-cases}

Although \toolname is general in its design and scope of applicability, in the rest of the paper, we will consider a use case of practical relevance in O-\ac{ran} systems. Specifically, we consider~\cite{tmc-coloran-wines} where the authors developed a set of xApps jointly controlling \ac{ran} slicing and scheduling policies for a set $\mathcal{L}$ of slices: \textit{(i)} \ac{embb}; \textit{(ii)} \ac{mmtc}; and \textit{(iii)} \ac{urllc}. We use the configurations and xApps embedding the pre-trained agents, which were provided upon request by the authors. Each agent optimizes resource allocation policies for an Open \ac{ran} \ac{gnb}. For each slice, the \ac{drl} agent selects a \ac{ran} slicing policy (\ie the number of \acp{prb} reserved to the slice), and the optimal scheduling policy to serve the \acp{ue} of the slice among Round Robin (RR), Waterfilling (WF), and Proportional Fair (PF).

\begin{figure}[tbp]
\centering%
\includegraphics[width=.75\textwidth,keepaspectratio]{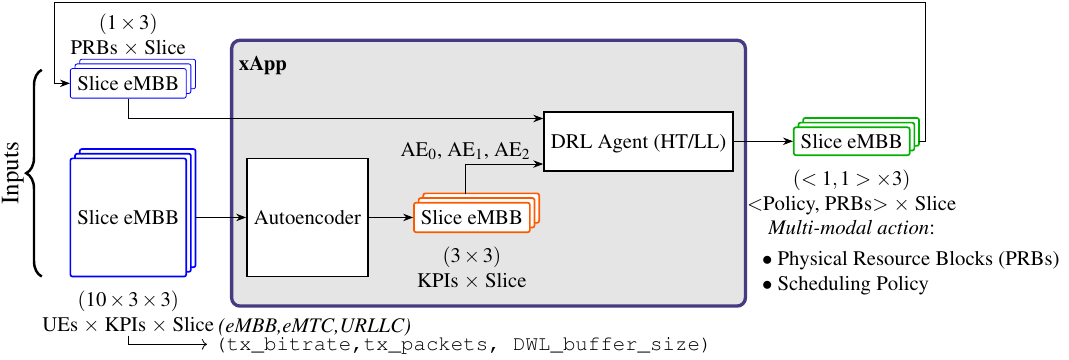}%
\vspace*{-2ex}%
\caption{The {\ac{drl}} framework consisting of an autoencoder and a \ac{drl} agent}%
\label{fig:sketch-architecture}%
\vspace*{-5ex}%
\end{figure}

The \ac{drl} agent depicted in Figure~\ref{fig:sketch-architecture} takes actions to maximize a target reward by monitoring a set $\mathcal{K}$ of \acp{kpi} received from the \ac{gnb} via the E2 interface and processed by an autoencoder. Specifically, the input $\mathbf{I}$ to the xApp consists of a $M\times K\times L$ matrix where $L=|\mathcal{L}|=3$ represents the number of slices, $K=|\mathcal{K}|=3$ is the number of monitored \acp{kpi} (\ie transmission bitrate in Mbps, number of transmitted packets, and size of the \ac{dwl} buffer in bytes), and $M=10$ is the number of individual measurements collected over the E2 interface for each slice. With a slight abuse of notation, let $\mathcal{K}=\{\txbrate, \txpkts, \dlbuff\}$ be the set of input \acp{kpi}. The generic element of the input matrix $\mathbf{I}$ is denoted as $i_{m,k,l}$ with $m=1,\dots,M$,  $k\in\mathcal{K}$, and $l\in\mathcal{L}$.

$\mathbf{I}$ is fed to an autoencoder that produces a latent representation of the input of size $K \times L$ (AE$_0$, AE$_1$, AE$_2$ in Figure~\ref{fig:sketch-architecture}). This low-dimensional representation is then fed to the \ac{drl} agent which embeds a \ac{ppo} architecture to compute a $c$-mode action (\ie $c=2$ in our use case) representing the combination of per-slice scheduling and slicing policies. At each time step $t$, the agents compute a reward function that maximizes the weighted sum of average \ac{kpi} values for each slice:
\begin{align}%
r_t(\mathbf{I}) = \sum_{l\in\mathcal{L}} w_l \cdot \sum_{m=1}^M \frac{i_{m,\kappa(s),l}}{M},%
\label{eq:reward}
\end{align}
where $\mathbf{I}$ represents the $M\times K\times L$ \ac{kpi} input matrix, and $\kappa(s)\in\mathcal{K}$ is used to indicate and extract only the target \ac{kpi} for each slice from $\mathbf{I}$. Specifically, for the \ac{embb} slice the target \ac{kpi} is \txbrate, while the target \acp{kpi} for the \ac{mmtc} and \ac{urllc} slices are \txpkts and \dlbuff, respectively. $w_l$ is a real-valued parameter used to weight the importance of each slice toward the reward maximization goal. $w_l$ takes positive values to maximize the reference \acp{kpi} of \ac{embb} and \ac{mmtc} slices and is negative to minimize \dlbuff for the \ac{urllc} slice as a proxy for minimizing latency. We extend~\cite{tmc-coloran-wines} and consider two \ac{drl} agent configurations:\footnote{We only retain basic configurations like scaling and normalizing the \acp{kpi} in the range $[-1,1]$.}\par

\noindent $\bullet$ \textit{\ac{ht}} prioritizes \ac{embb} slice's reward contribution over the other two;\par
\noindent $\bullet$ \textit{\ac{ll}} prioritizes the contribution of the \ac{urllc} slice over the other two slices.

These agents control the \ac{ran} which is a highly non-stationary and dynamic environment with changing channel conditions and they handle diverse traffic profiles in each slice. There are obvious tradeoffs behind the agents' decisions: assigning too many \acp{prb} to the \ac{embb} slice substantially reduces the throughput that \acp{ue} of the other two slices will experience (see \S~\ref{sec:experiments}). Furthermore, agents deal with a continuous state space, the output of the autoencoder, which also makes it hard to quantify the number of states of the system. Therefore, it becomes essential for domain experts access \ac{xai} tools to better comprehend agents' decisions under time-varying network conditions.

\subsection{Applying XAI Tools Out-of-the-Box}
\label{subsec:motivation-shap}
We now elaborate the need for \toolname by showing the inherent limitations of popular \ac{xai} tools (\ie \ac{shap} and \acp{dt}) applied to the \ac{ht} and \ac{ll} agents described in \S~\ref{subsec:drl-use-cases}.

\ac{shap} has been proven effective in a number of scenarios in combination with \ac{drl}-based solutions. Recent works have leveraged \ac{shap} for power- and \ac{uav}-control~\cite{zhang-shap-drl-power-tcss,HE2021107052}, biomedical scenarios~\cite{shap-biomedic} and in the presence of multi-agent systems~\cite{alexandre-shap-coll-multiagent-drl-mag,li-shap-multiagent-drl-kdd}. In a nutshell, \ac{shap} provides feature-based interpretations by approximating the Shapley values of a prediction and generates global and local explanations in the form of log-odds, which can be turned into a probability distribution with the \texttt{softmax} operation. 

\acp{dt} can also explain \ac{drl} agents. Metis~\cite{zili-interpret-sigcomm20} converts \ac{dnn} and \ac{drl} solutions into interpretable rule-based configurations via \acp{dt} and hypergraphs. Trustee~\cite{trustee} constructs a high-fidelity and intuitive \ac{dt} taking as input both the \ac{ai} model and training dataset. If applied out-of-the-box, \ac{shap} and \acp{dt} would be applicable only at \ac{drl} agent block of Figure~\ref{fig:sketch-architecture}. We now elaborate why both are ill-suited to deliver interpretations on the whole architecture.

\begin{figure*}
\captionsetup[subfigure]{captionskip=-2pt}
\centering
\hspace*{2.1em}\includegraphics[scale=.5]{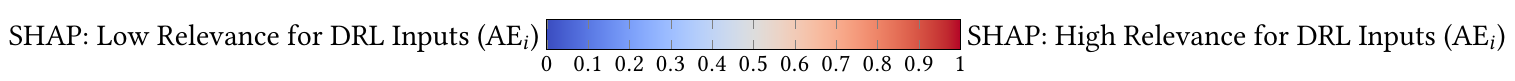}\par
\vspace*{-3.5ex}%
\subfloat[\ac{embb} slice~\label{fig:shap-sl0}]{%
    \includegraphics[width=.32\textwidth,keepaspectratio]{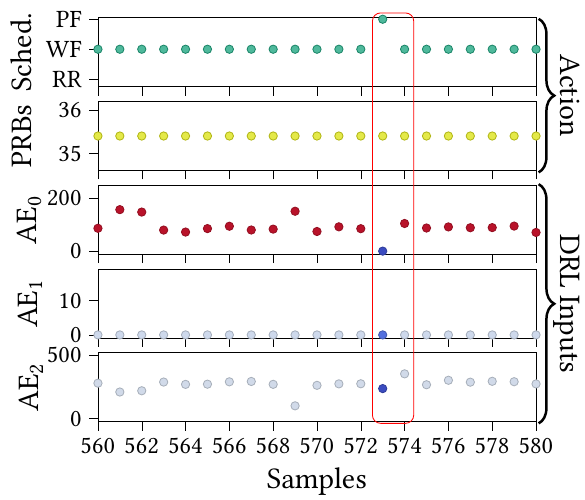}}%
~%
\subfloat[\ac{mmtc} slice~\label{fig:shap-sl1}]{%
    \includegraphics[width=.307\textwidth,keepaspectratio]{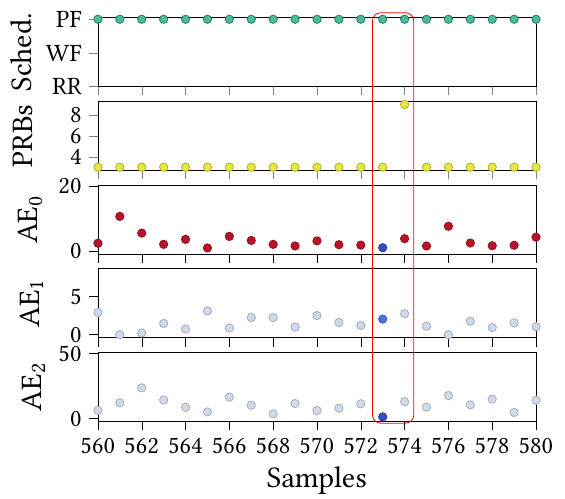}}%
~%
\subfloat[\ac{urllc} slice~\label{fig:shap-sl2}]{%
    \includegraphics[width=.307\textwidth,keepaspectratio]{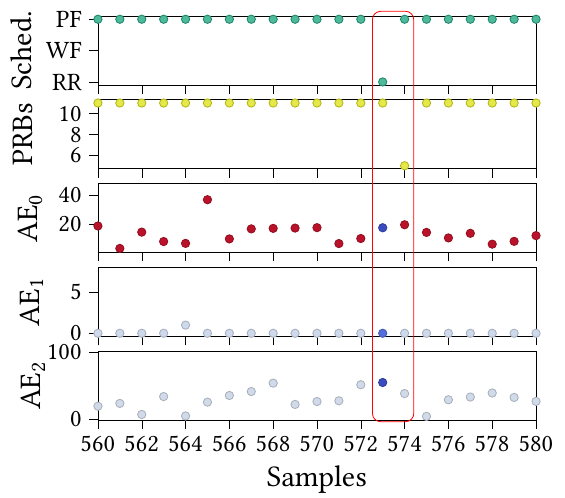}}%
\vspace*{-1.5ex}%
\caption{The fine-grained details of {\ac{shap}} explanations for the \ac{ht} agent}%
\label{fig:shap-motivation}%
\vspace*{-4ex}%
\end{figure*}

Figure~\ref{fig:shap-motivation} portrays an example of the outcome of applying \ac{shap} to the \ac{ht} agent. For each slice, \ac{shap} computes relevance scores by determining the average contribution of each element of the input across all possible permutations of elements' values with respect to the output of the agent, i.e., the corresponding action, expressed as the number of \acp{prb} (taking values in $[0,50]$) and scheduling policy (among RR, WF, PF) assigned to the slice. The inputs of the \ac{drl} agent are the outputs of the autoencoder (AE$_0$, AE$_1$, and AE$_2$ in Figure~\ref{fig:sketch-architecture}) and not the actual \emph{inputs} of the system, \ie the \acp{kpi}. Formally $\forall i = 1, 2, \ldots, N$, with $N=K \times L$, the score $r_i \in R_N$ is computed as:
\begin{equation}\label{eq:shap-scores}
r_i(f) = \frac{1}{(N-1)!} \sum_{k=1}^{N-1} \sum_{\substack{X_s \subseteq X_t}, |s|=k} \left[ \binom{N-1}{k} \right]^{-1}
\cdot \left( f(X_t) - f(X_s) \right),
\end{equation}
where $f(X_t)$ is the action taken considering all the features $X_t=\{\mathrm{AE}_0, \mathrm{AE}_1, \mathrm{AE}_2\}$, $s=N-1$ is a subset of the $N$ features of the input sequence, and $f(X_s)$ is the action taken under input $X_s$.

We now elaborate on the pros and cons of \ac{shap} as a result from extensive tests executed for both \ac{ht} and \ac{ll} agents in various settings that include varying the number of users per slice and traffic scenarios (see~Table~\ref{tab:exp-settings} in Appendix~\ref{app-sec:exp-config}). \ac{shap} is extremely precise in identifying which feature is the most important for taking an action, and reveals the precise operation of the agent. Figure~\ref{fig:shap-motivation} shows for 20 time steps the values of the \ac{drl} inputs and outputs. The colors of the color-bar identify the relevance scores of the \ac{drl} inputs computed with \ac{shap}: blue and red colors correspond to low and high scores respectively. Samples denoted with low relevance in all the \ac{drl} inputs (see index $573$ in Figure~\ref{fig:shap-motivation}) trigger a change in scheduling policy which follows a change in \ac{prb} allocation (\eg the scheduling policy of the \ac{embb} slice transitions from WF to PF - highlighted with frames in Figure~\ref{fig:shap-motivation}). However, \ac{shap} \textit{(i)} is limited by the autoencoder to show non-intuitive explanations, \ie feature relevance of autoencoder outputs per policy and for each slice and not the actual inputs (the \acp{kpi} at user level); and \textit{(ii)} is extremely costly from a computational perspective. Even for few users, computing \ac{shap} values on Nvidia RTX 3090 and A100 SXM4 GPU cards can take hours (see Figure~\ref{fig:shap-proc-requirements}(a)): this holds across agents (see Figure~\ref{fig:shap-proc-requirements}(b)) and for the other configurations tested. Note that increasing the number of users from 4 to 6 does not produce tangible changes.

\medskip
\begin{minipage}{0.65\textwidth}%
\captionsetup[subfigure]{captionskip=0pt}%
\captionsetup[sub]{labelformat=parens,labelsep=space}
\centering
\includegraphics[width=.49\textwidth,keepaspectratio]{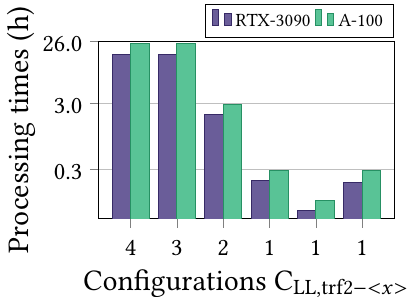}%
~%
\includegraphics[width=.49\textwidth,keepaspectratio]{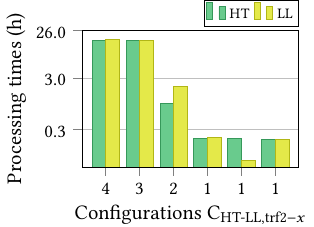}%
\par
{\centering \small\sffamily%
(a) On different machines%
\quad%
(b) On different agents%
}%
\vspace*{-2ex}%
\captionof{figure}{\ac{shap} computational complexity. Ticklabels on x-axis denote the number of $<$$x$$>$ users in the experiment. Experiments with 1 user only are executed for the three slices.}%
\label{fig:shap-proc-requirements}%
\end{minipage}
~%
\begin{minipage}{0.3\textwidth}%
\centering
\vspace*{-2ex}%
\captionof{table}{Classification accuracy}%
\label{tab:dt-accuracy}%
\vspace*{-2ex}%
\resizebox{0.85\textwidth}{!}{%
\begin{tabular}{r>{\RaggedRight}p{2.25cm}}%
\toprule
\textsc{Config.} & \textsc{\ac{dt} Accuracy}\\
\midrule
C$_{\text{\ac{ll},trf1}-4}$ & $18.74$\,\% \\
C$_{\text{\ac{ht},trf1}-3}$ & $43.35$\,\% \\
C$_{\text{\ac{ll},trf2}-3}$ & $58.52$\,\% \\
C$_{\text{\ac{ll},trf1}-1}$ & $23.20$\,\% \\
C$_{\text{\ac{ht},trf1}-1}$ & $35.71$\,\% \\
C$_{\text{\ac{ht},trf2}-1}$ & $37.86$\,\% \\
\bottomrule
\end{tabular}%
}%

\vphantom{\rule{2em}{9ex}}
\end{minipage}
\medskip

We also build \acp{dt} via an XGBoost model~\cite{chen2016xgboost} with $N$ features and target label equal to the action, i.e., the output of the \ac{drl} agent. Table~1 shows that the ensemble of decision trees is not performing well in the classification task across different configurations. This prevents from explaining how the agent takes an action upon observing $X_t$. As this part does not become interpretable, it is not possible to provide explanations to the overall framework of Figure~\ref{fig:sketch-architecture} via divide and conquer (i.e., explaining first the \ac{drl} agent and backpropagate to the input by explaining the autoencoder).

In conclusion, \ac{shap} is too fine-grained and slow. Both \ac{shap} and \acp{dt} are unable to provide intuitive explanations to link agent outputs, i.e., actions, with their impact on \acp{kpi}, e.g., \txbrate.

\subsection{XAI Challenges}
\label{subsec:challenges}

Based on the above discussion, we identify the following three major challenges:
\begin{enumerate}[label=$\bullet$ \textit{Challenge\,\arabic*:},
ref=\textit{Challenge\,\arabic*}, 
wide=0\parindent,
listparindent=0pt,
align=left]
\item \label{cha:1} While autoencoders are of great help to reduce the input dimension and facilitate generalization, they also eliminate the direct \mbox{\emph{input-output}} connection that state-of-the art \ac{xai} techniques such as \ac{shap} and \ac{lrp} build upon. That is, when using autoencoders, \ac{shap} and \ac{lrp} can only reveal the contribution that the latent space representation had on the action taken by the agent, but lose any information on how the actual input affected the decision-making process.
\item \label{cha:2} Whenever actions depend on either \textit{past} actions or states of the environment (\eg previous \ac{prb} allocation), the decision-making process relies on memory. Such feedback loop adds an additional layer of complexity to the already complex and non-linear relationship between inputs and outputs. This prevents the use of well-established tools such as casual models~\cite{madumal-xlr-casual-lens} as it becomes harder to identify the direction of causality, primary cause and primary effect of an action. 
\item \label{cha:3} Unlike many popular \ac{drl}-based agents that control individual parameters~\cite{mao-pensieve-sigcomm17,tarik-drl-routing-22}, actions that agents take in O-RAN systems are likely multi-modal and involves several control parameters at the same time like \ac{ran} slicing and scheduling policy. This makes it hard to leverage existing \ac{xai} tools which are primarily tailored to explaining much simpler control actions. 
\end{enumerate}

With \toolname, we aim to address these challenges and provide explainability for the class of \ac{drl}-based Open RAN solutions described above. Specifically, we seek explanations that are intuitive, \eg which actions determine changes in \acp{kpi}. We believe that a clear understanding and programmability of the agent behavior is key to lower the barrier for adopting \ac{drl} in operational O-RAN networks and provide operators with the necessary tools to understand why the AI has taken a certain decision thus building useful knowledge to design and deploy more efficient networks.

\section{The \toolname Framework}
\label{sec:interpretation}

Because of above-mentioned challenges, we deal with systems where the \emph{input}-\emph{output} link is not available (\ref{cha:1}), the decision making process relies on memory (\ref{cha:2}), and actions are multi-modal (\ref{cha:3}). We address these challenges by embedding into attributed graphs multi-modal actions (nodes) and their impact on the future state (attributes). This allows recording the \emph{consequence} of an action. To distill knowledge, we analyze transitions over time (edges) and quantify statistically the distance between the attributes of the respective nodes. Thus, we can explain the agent behavior by determining the effect of its decisions on the environment with details on the contribution of each component of the multi-modal action.

Next, we first motivate the choice of attributed graphs and elaborate why other data structures like \acp{dt}, hypergraphs or multi-layer graphs are not effective (\S~\ref{subsec:explore-design}), describe the \toolname architecture and elaborate on how it interacts with \ac{drl} agents (\S~\ref{subsec:explore-interaction-drl-agent}). Then, we explain how to distill knowledge from the attributed graph (\S~\ref{subsec:explore-synthesis}). Finally, we show how to leverage the explanations to improve the \ac{drl} agent's decision-making process (\S~\ref{subsec:explora-optimize}).

\subsection{Design Choices}
\label{subsec:explore-design}
Domain experts seek to receive from \ac{xai} tools explanations inherently related to understanding the logic and dynamics that tie \emph{inputs} to \emph{outputs}~\cite{zili-interpret-sigcomm20}. For \ac{drl}, this directly translates into understanding the reason why an agent has taken action $a_t$ at time $t$ when observing state during a window $\delta$ before $t$, \ie $s_{(t-1)+\delta}$. Deriving such logic is not easy, as the unidirectional link \mbox{\emph{input-output}} is essentially broken by the introduction of the autoencoder (\ref{cha:1}). However, we leverage the fact that any action $a_t$ will alter the future state $s_{t+\delta}$, which is observable even with the autoencoder. Next, we generate knowledge by analyzing the changes in subsequent actions $a_t\rightarrow a_{t+1}$ and the corresponding change of states $s_{t+\delta}\rightarrow s_{t+1+\delta}$. We thus seek a data structure capable of capturing such connections.

In the past, explainability has been delivered via several data structures~\cite{du2019techniques,guidotti2018survey}, with the most well-established being \acp{dt}~\cite{moulay2022automated,zili-interpret-sigcomm20} and hypergraphs~\cite{zili-interpret-sigcomm20}. For example, Metis~\cite{zili-interpret-sigcomm20} combines both \acp{dt} and hypergraphs to distill knowledge and has proved successful when applied to Pensieve~\cite{mao-pensieve-sigcomm17}, an Adaptive Bit Rate (ABR) \ac{drl} system that optimizes video bitrate selection (\ie the action) by observing and adapting to past video chunk bitrate, throughput, buffer occupancy (\ie the state). However, \ac{xai} tools that use \acp{dt} and hypergraphs can be applied only to \ac{drl} agents like Pensieve that take unimodal actions such as selecting the bitrate on a per-user basis, which makes them unsuitable to address both \ref{cha:1} and \ref{cha:3}. When applying \acp{dt} to \ac{drl} agents with multi-modal actions, we observed a lack of generalization resulting from attempts of pruning the excessive growth scale of the \acp{dt}.

We resort to attributed graphs~\cite{lawrence2013attributed}. Formally, an attributed graph $G$ is defined as $G=(N,E,B)$ where $\mathcal{N}$ is a set of nodes, $\mathcal{E} \subseteq \mathcal{N} \times \mathcal{N}$ is a set of edges, and $\mathcal{B}$ is a set of attributes associated with $\mathcal{N}$. Specifically, for each node $n\in \mathcal{N}$, there exists an attribute $b(n)\in \mathcal{B}$ consisting of at most $P$ elements: $b(n)=\{b_1(n),\ldots,b_p(n)\}$. Multi-layered graphs and hypergraphs are useful mathematical tools to model complex and multiple relations among multiple entities, like resource allocation in cloud-\ac{ran} systems where users are connected to radio units depending on channel conditions and radio units are connected to centralized units according to traffic load~\cite{zhang-hypergraphs-5g}. In contrast, attributed graphs capture effectively individual relationships between entities against a common set of properties. This is precisely the case of \ac{drl} agents where entities are actions, the relation between entities is temporal (i.e., action $a_{t+1}$ occurs after $a_t$) and the common set of properties of the entities maps to the state $s$ associated to each action $a$.
We will describe how to build the attributed graph in \S~\ref{subsec:explore-interaction-drl-agent}.

\subsection{\toolname: System Architecture}
\label{subsec:explore-interaction-drl-agent}

\begin{figure}[tb]
\centering%
\includegraphics[width=.55\textwidth, keepaspectratio]{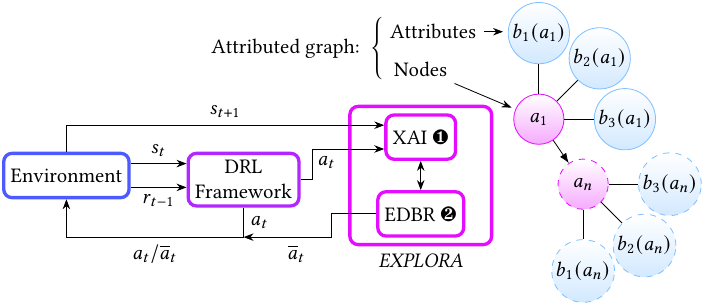}%
\vspace*{-2ex}%
\caption{\toolname interaction with a \ac{drl} agent}%
\label{fig:explore-interaction-drl}%
\vspace*{-4ex}%
\end{figure}

\simpletitle{The architecture} \toolname consists of the \textit{\ac{xai} module} (\ding{182} in Figure~\ref{fig:explore-interaction-drl}) and the \ac{xai}-aided \textit{Explanation-Driven Behavior Refiner (EDBR) module} (\ding{183} in Figure~\ref{fig:explore-interaction-drl}). \ding{182} generates post-hoc explanations about the agent behavior by building the attributed graph $G$ throughout an observation window $W$. \ding{183} leverages such explanations to understand the decision-making process (\S~\ref{subsec:explore-synthesis}), identifies inefficiencies and improves overall network performance (\S~\ref{subsec:explora-optimize}).

Figure~\ref{fig:explore-interaction-drl} shows how these two blocks interface with the \ac{drl} agent as well as their workflow execution. The \ac{drl} agent interacts with the environment as described in \S~\ref{subsec:bck-drl}, with the sole exception that the input is the latent representation and not the state. This approach can be easily applied to a variety of \ac{drl} models such as \ac{dqn}~\cite{tarik-drl-routing-22}, \ac{ppo}~\cite{tmc-coloran-wines} or \ac{a3c}~\cite{mao-pensieve-sigcomm17}. These differ in the way the agent learns the optimal policy.

\simpletitle{Generating the attributed graph} To generate the attributed graph $G$ described in the previous section, \toolname's \ac{xai} module (\ding{182} in Figure~\ref{fig:explore-interaction-drl}) performs the following operations. During $t\in W$, each action $a_t=(a_t^1,\dots,a_t^c)$ (representing the decision of \ac{ran} slicing and scheduling policies with $c=2$ in our use case) is mapped to a node $n$ of the graph. The impact of $a_t$ on the future environment state, $s_{t+1}$, is instead mapped to the attribute $b(n)$. For our use case, we embed the distribution for each monitored \acp{kpi} and for each slice as the $b_p(n)$ element of the attribute, where $p \in \mathcal{P}$, $\mathcal{P}=\mathcal{K}\times \mathcal{L}$ and $P=K\cdot L$. From \S~\ref{subsec:drl-use-cases} and \S~\ref{subsec:bck-drl}, the action $a \in \mathcal{A}$ is multi-modal and can be defined as a $c$-tuple \mbox{$a=(a^1,\ldots,a^c)$}. Similarly, the state $s_t \in \mathcal{S}$ can be defined as a tuple which, in our case, is the input matrix $\mathbf{I}$. An edge connecting actions $a_t$ and $a_{t+1}$ ($a_1$ and $a_{n}$ in Figure~\ref{fig:explore-interaction-drl}) indicates the transition between actions taken at subsequent time instances. As $b(a_t)$ indicates the effect of $a_t$ on $s_{t+1}$, then the unidirectional edge $a_t \rightarrow a_{t+1}$ indicates that $b(a_t)$ is the input state space for $a_{t+1}$. This closes the loop broken by the presence of the autoencoder. \acp{dt} and hypergraphs fail precisely in providing a link connecting actions to their consequences onto the environment and then on the next action. The same would apply to multi-layer graphs embedding one of the components of the multi-modal action in each layer. After $W$ observations, $G$ makes it possible to distill knowledge regarding the agent's behavior (\S~\ref{subsec:explore-synthesis}) and the expected outcome of an action is known in advance and before it is actually enforced.

In the majority of \ac{drl}-based systems 
for Open RAN, the state and action spaces $\mathcal{S}$ and $\mathcal{A}$ of the \ac{drl} agents might be very large, as they embed real-valued variables such as throughput, channel conditions, buffer size, latency, to name a few. Therefore, to tackle this complexity and contain the size of the attribute space, we operate as follows. First, the set $\mathcal{P}$ has limited dimension, as it depends on the number $K$ of \acp{kpi} included in the state $s$ and the number $L$ of slices. Second, each element $b_p(n)\in b(n)$ only stores the distribution of each \ac{kpi}. 
In the right portion of Figure~\ref{fig:explore-interaction-drl}, we show an example of the attributed graph $G$ with two actions (\ie nodes) $a_1$ and $a_n$, each storing $3$ attributes (i.e, $b_1(a_1)$, $b_2(a_1)$, and $b_3(a_1)$ for $a_1$). In Appendix~\ref{app-sec:example-graph}, we provide explanations about the generation of $G$ for three consecutive step. 
Next, we show how to distill knowledge from $G$ (\S~\ref{subsec:explore-synthesis}), and how this knowledge can be used by module \ding{183} to improve overall performance (\S~\ref{subsec:explora-optimize}).

\subsection{Synthesizing Network-Oriented Explanations}
\label{subsec:explore-synthesis}

\toolname distills knowledge by analyzing transitions between actions (edges in $G$). This allows to characterize the individual contribution of each component of a multi-modal action. For example, in our use case, the graph will reveal if a multi-modal action (recall that $c=2$) produces changes in \acp{kpi} like throughput because of a change in \ac{ran} slicing, scheduling policies or both. This would not be possible with other data structures that do not link attributes (\acp{kpi}) to nodes (actions). Any action taken at time $t$ can be expressed as a $2$-tuple, (in our case, slicing and scheduling policy) $a_t=(a_t^1,a_t^2)$. Then, at any transition between $t$ and $t+1$, there exists $2^c$ possible combinations to describe how actions transition between $a_t$ and $a_{t+1}$. For example, in one case $a_t^1=a_{t+1}^1$ but $a_t^2\neq a_{t+1}^2$. In another, $a_t^1\neq a_{t+1}^1$ and $a_t^2 = a_{t+1}^2$. The remaining two cases are: the case where the action remains the same, \ie $a_t=a_{t+1}$, and the one where they are completely different, \ie $a_t^i\neq a_{t+1}^i$ for $i=1,2$.

\toolname builds a set of ordered pairs $(\pi, v)$, where each $\pi$ maps the action transition $a_t \rightarrow a_{t+1}$ and $v$ maps the corresponding change of impact to the respective states $s_{t+1} \rightarrow s_{t+2}$ (in our use case, each $\pi$ would correspond to a \ac{ran} slicing and policy scheduling enforced in two subsequent timesteps). We can now quantify such impact via the attributes $b(n)$ and $b(n+1)$ of the nodes $n$ and $n+1$ representing the action transition $a_t \rightarrow a_{t+1}$. Recalling that each $b_p(n)$, with $p \in \mathcal{P}$ embeds a distribution of each \ac{kpi} for each slice, we can compare each $b_p(n+1)\rightarrow b_p(n+2)$ using either statistical techniques like the Jensen Shannon divergence or, for example, directly comparing $\avg\{b_p(n+1)\}$ and $\avg\{b_p(n+1)\}$. The result of such comparison is stored in $v={v_1, v_2, \ldots, v_p}$ and is the knowledge we leverage to produce informative explanations, \ie which is the effect that changes of \ac{ran} slicing and/or scheduling policies produce on \acp{kpi} like throughput or buffer size. To distill knowledge, \toolname uses \acp{dt}. Specifically, it builds a \ac{dt} where the set of features is $v$ and the target label is the corresponding class of action transition among the $2^c$ possible combinations. The resulting \ac{dt} identifies which changes on \acp{kpi} are associated to each class of transitions and the visual inspection of the tree provides informative explanations in the form: \textit{``the agent uses completely different transitions ($a_t^i\neq a_{t+1}^i$) to increase the \txbrate}.'' Note that the use of \acp{dt} for knowledge distillation does not imply that \acp{dt} could substitute the original \ac{drl} agents in taking joint decisions on \acp{prb} allocation and scheduling policy. We will show in \S~\ref{subsec:results-explanations} how to turn these explanations into a concise and effective summary of the agent's behavior.

\subsection{Optimizations Enabled by \toolname}
\label{subsec:explora-optimize}
With the distilled knowledge generated by \ding{182}, \toolname's EDBR module (\ding{183}) makes it possible to perform informed and targeted ad-hoc adjustments to the agent's behavior to modify its decision-making process with the goal of improving the overall network performance. This fits well with intent-based networking~\cite{aris-ibn-comst} which aims at delivering a simplified and agile network management where complex configurations are translated into high-level intents.

Following O-RAN specifications, \ac{drl} agents are usually trained offline\footnote{Note that this is a strict requirement for any AI-based xApp and rApp~\cite{comst-surveoran-wines}.} on data that might not necessarily accurately reflect the type of data observed in real-time. This is because an ideal training is extremely complex\footnote{To train agents \ac{ht} and \ac{ll}, the data collection took two and a half months on Colosseum and the actual training operation took approximately $10$-$1$5 hours on a single NVIDIA A100 GPU, depending on the model.} and impractical for production networks where the number of scenarios to be accounted for is huge (e.g., number of served users, traffic dynamics, propagation characteristics of the environment, among others). Optimal actions on training data may thus perform poorly in a live network, \eg because of different traffic profiles and topologies~\cite{tmc-coloran-wines}. 
\toolname builds and updates $G$ over time, and hence can be used to identify inefficient actions that are attributed to imperfect training and offer mechanisms that prevent inefficiencies. While the explanations can simply be used to understand how agents make decisions, we further discuss two concrete possible uses that domain experts can make of the distilled knowledge.
\begin{enumerate}[label=$\bullet$ \textit{Opt\,\arabic*:},
ref=\textit{Opt\,\arabic*}, 
wide=0\parindent,
listparindent=0pt,
align=left]
\item \label{opt:1} \textit{Intent-based action steering} replaces an action selected by the agent with another extracted from the attributed graph $G$ to fulfill specific intents. For example, if an agent takes an action that $G$ marks as potentially resulting in an expected low reward, \toolname can suggest an alternative action from $G$ that would yield a higher expected reward.
\item \label{opt:2} \textit{Action shielding} prevents the agent from taking specific actions considered dangerous. Unlike action steering, action shielding completely inhibits specific actions and is mainly used for security purposes~\cite{alshiekh-shield-drl} (\eg for active voltage control~\cite{chen-shield-example-drl}).
\end{enumerate}

Given the highly non-stationary dynamics of the \ac{ran} and the class of resource allocation \ac{drl}-based solutions considered in this work, \ref{opt:1} looks more attractive as it enables to programmatically control the agent behavior \emph{with consciousness} and thanks to the understanding of the impact that certain actions have on the state of the network. We will discuss policies for action steering in \S~\ref{subsec:impl-ar-strategies}, and show such benefits in \S~\ref{subsec:results-optimization}.

\section{\toolname Implementation}
\label{sec:implementation}
In this section, we illustrate how to embed \toolname into xApps (\S~\ref{subsec:impl-packaging}), and elaborate on possible action replacement strategies to improve overall \ac{kpi} performance (\S~\ref{subsec:impl-ar-strategies}).

\subsection{Integrating \toolname in xApps}
\label{subsec:impl-packaging}

\begin{figure}[tb]
\centering%
\includegraphics[width=.6\textwidth, keepaspectratio]{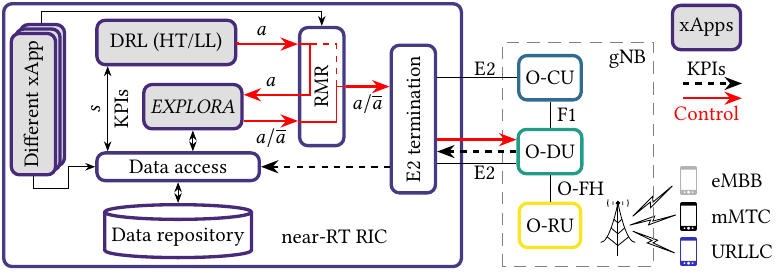}%
\vspace*{-1.5ex}%
\caption{\toolname integration into the \oran reference architecture}%
\label{fig:explore-in-oran-arch}%
\vspace*{-4ex}%
\end{figure}

Thanks to the flexibility offered by the O-RAN architecture, we identify three possible strategies to make \toolname an operational component of the near-RT \ac{ric}. Specifically, it can be: \textit{(i)} a base component of the \ac{ric} itself hosted outside the xApp domain; \textit{(ii)} a component of each xApp, embedded in the microservice that executes the \ac{drl} agent; or \textit{(iii)} a standalone xApp that interacts with one or more xApps hosting \ac{drl} agents. Strategy \textit{(iii)} provides the best trade-off between flexibility and cost: the dedicated \toolname xApp can be replicated as needed to support diverse use cases and interacts directly with the xApps hosting \ac{drl} agents without requiring changes as in strategy \textit{(ii)}, or to the near-RT {\ac{ric}} platform.

The left part of Figure~\ref{fig:explore-in-oran-arch} illustrates how the \toolname xApp interacts with other components of the O-RAN architecture. The E2 termination of the near-RT \ac{ric} platform routes E2 data (\acp{kpi}) to a data access microservice that stores it in a data repository inside the \ac{ric}. The same microservice is queried by the \ac{drl} and \toolname xApps to access stored data. As discussed in \S~\ref{subsec:drl-use-cases}, the \ac{drl} xApp uses the \acp{kpi} to first feed the autoencoder and then the agent (\ac{ht} or \ac{ll} in our use case), thus making such \acp{kpi} represent the state $s$ observed by both the \ac{drl} and \toolname xApps.

The \ac{drl} xApp then generates an action $a$ in a RIC internal message (\eg for the E2 service model \ac{ran} Control) that is sent to the \ac{rmr}, as shown in Figure~\ref{fig:explore-in-oran-arch}. The \ac{rmr} is in charge of dispatching internal messages across different components, \eg xApps and E2 termination, and can be configured with ad-hoc routes based on the endpoints and message ID~\cite{oran-wg3-ricarch}. Therefore, to seamlessly integrate the \toolname xApp within the RIC, we configure the \ac{rmr} to route the RAN control messages to the \toolname xApp. If the \toolname xApp is not deployed, the \ac{rmr} delivers the control message to be enforced at the O-DU from the \ac{drl} agent xApp via the E2 termination directly (see the red dashed line in Figure~\ref{fig:explore-in-oran-arch}).

Once the \toolname xApp receives action $a$, it can decide whether to forward it \textit{as is} to the \ac{rmr}, which then sends it to the E2 termination, or to update it with an action $\overline{a}$ computed by following one of the strategies discussed in \S~\ref{subsec:impl-ar-strategies}. The \toolname xApp can also interact with the data access microservice to save the explanations and information on the state/action/explanation tuple in the data repository. This can later be accessed by the network operator for quality assurance, debugging, and/or datasets generation purposes. 

\subsection{Strategies for Action Steering}
\label{subsec:impl-ar-strategies}

We now show ad-hoc adjustments that domain experts (\eg network operators) can use to ultimately improve users' \acp{kpi} by defining high-level intents. In the case of imperfect training (see \S~\ref{subsec:explora-optimize}), the agent might observe previously unseen input data, which might result in taking a sub-optimal (or inefficient) action $a_t$. $a_t$ is sub-optimal if its expected reward could be improved by another action $a_t^{'}$. The EDBR module uses the knowledge built and distilled via the synthesis of explanations from $G$ to suggest another action $\overline{a}_t$ whose expected reward and impact on the future state is statistically known from the past.
Algorithm~\ref{algo:strategies} provides the details of the implemented intent-based action steering strategies, which are summarized hereafter. For the sake of demonstration, and due to space limitations, we limit the graph exploration to the first hop nodes of $G$ only. This restriction highlights the benefits of the strategies in a worst-case scenario. 

\begin{enumerate}[label=$\bullet$ \textit{AR\,\arabic*~-},
ref=\textit{AR\,\arabic*}, 
wide=0\parindent,
listparindent=0pt,
align=left]
\item \label{ar:1} ``Max-reward'': this strategy replaces an action $a_t$ suggested by the agent that is expected to yield a low reward with one extracted from the graph ($a_G$) that is expected to yield a higher reward. This can be achieved by extracting the attributes $b(a_t)$ and $b(a_G)$ from the attributed graph $G$, and computing the expected reward (defined in \eqref{eq:reward}) using the average \ac{kpi} values stored in $G$. For the \acf{ht} agent, we expect this strategy to favor the \ac{embb} slice.
\item \label{ar:2} ``Min-reward'': this strategy substitutes an action $a_t$ suggested by the agent and expected to result in a high reward with an action $a_G$ from $G$ that is expected to yield a lower reward. For the \acf{ll} agent, we expect this strategy to favor the \ac{urllc} slice.
\item \label{ar:3} ``Improve bitrate'': similarly to ``Max-reward'', this strategy replaces an action $a_t$ computed by the \ac{drl} agent with an expected low reward with another action $a_G$ that is expected to deliver a high \txbrate. We expect this strategy to always favor the \ac{embb} slice for any agent.
\end{enumerate} 
\vspace*{-1ex}%

\begin{algorithm}%
\setlength{\textfloatsep}{-1ex}
\small
\caption{Strategies for intent-based action steering}%
\label{algo:strategies}%
\begin{algorithmic}[1]
\Require{$G=(N,E,B)$; action suggested by the agent $a_t$; previous action $a_{t-1}$; $O$; Steering strategy $\alpha \in \{\ref{ar:1}, \ref{ar:2}, \ref{ar:3}\}$}
\State $\omega \leftarrow$ Result of $r(b(a_t)) < \avg_{x=t-O-1}^{t-1}r(a_x)$
\If{$(\omega,\alpha)==(\texttt{True},\ref{ar:1})$ OR $(\omega,\alpha)==(\texttt{False},\ref{ar:2})$ OR $(\omega,\alpha)==(\texttt{True},\ref{ar:3})$}
\If{$n_{t-1} \in N$}
\State Initialize $Q\leftarrow \emptyset$
\State Mark $n_{t-1}$ as visited, add $n_{t-1},b(a_{t-1})$ to $Q$
\For{each neighbor $w$ of $n_{t-1}$}
\If{$w$ is not visited}
\State Mark node $w$ as visited
\State $a \leftarrow $ Action corresponding to node $w$
\State Add $w,b(a)$ to $Q$
\EndIf
\EndFor
\State Execute procedure $\alpha\in \{\ref{ar:1}, \ref{ar:2}, \ref{ar:3}\}$
\Else
\State Send $a_t$ to RMR
\EndIf
\EndIf
\vspace{0.1cm}
\Procedure{\ref{ar:1}: ``Max-reward''}{$Q$,$a_t$}
\State $a_{max} = \arg\max_{a} \{r(b(a)) : b(a) \in (w,b(a)) \in Q\}$
\If{$r(b(a_{max})) > r(b(a_t))$}
\State $a_t \leftarrow$ $a_{max}$
\EndIf
\State Send $a_t$ to RMR
\EndProcedure
\vspace{0.1cm}
\Procedure{\ref{ar:2}: ``Min-reward''}{$Q$,$a_t$}
\State $a_{min} = \arg\min_{a} \{r(b(a)) : b(a) \in (w,b(a)) \in Q\}$
\If{$r(b(a_{min})) < r(b(a_t))$}
\State $a_t \leftarrow$ $a_{min}$
\EndIf
\State Send $a_t$ to RMR
\EndProcedure
\vspace{0.1cm}
\Procedure{\ref{ar:3}: ``Improve bitrate''}{$Q$,$a_t$}
\State $j \leftarrow $ Index of \txbrate KPI in the attributes $b(\cdot) \in G$
\State $a_{br} = \arg\max_{a} \{b_j(a)) : b_j(a) \in b(a) \in (w,b(a)) \in Q\}$
\If{$b_j(a_{br})) > b_j(a_t))$}
\State $a_t \leftarrow$ $a_{br}$
\EndIf
\State Send $a_t$ to RMR
\EndProcedure
\end{algorithmic}%
\vspace*{-1ex}%
\end{algorithm}
With a slight abuse of notation, let $r(b(a_t))$ be the reward computed from \eqref{eq:reward} where instantaneous \acp{kpi} are replaced with their average values computed from the distributions stored in the attributes $b(a_t)$ for the action $a_t$. For all the strategies, we compare the expected reward $r(b(a_t))$ that would be achieved by taking action $a_t$ with the measured average reward $\avg_{x=t-O-1}^{t-1}r(a_x)$ we have observed across the last $O$ time steps.

\section{Experimental Evaluation}
\label{sec:experiments}

In this section, we first describe the \oran platform used to validate \toolname in a broad range of settings (\S~\ref{subsec:results-settings}). Then, we empirically evaluate \toolname explanations (\S~\ref{subsec:results-explanations}) and benchmark its ability to programmatically improve the agents' behavior with ad-hoc adjustments (\S~\ref{subsec:results-optimization}). 

\subsection{\oran Testbed and Setup Description}
\label{subsec:results-settings}
We leverage the open-source OpenRAN Gym framework~\cite{openrangym} to deploy a reference O-RAN architecture (as that of Figure~\ref{fig:explore-in-oran-arch}) with our custom xApps and perform data collection. If features an O-RAN-compliant \mbox{near-RT \ac{ric}}~\cite{amber_release}; instances of the E2 interface to connect the RIC and the RAN; integrates \acp{gnb} and \acp{ue} from open-source software-defined 3GPP stacks (specifically, we used srsRAN \acp{bs} and \acp{ue} for this study); and features stubs for xApps that can be extended to implement the desired control functionalities, \ie \toolname and the \ac{drl} agents.

We deploy OpenRAN Gym components in Colosseum, a wireless emulation testbed with hardware in the loop~\cite{colosseum}. Colosseum provides 128 pairs of programmable compute nodes with a white-box server and a software-defined radio (NI/Ettus USRP X310). These nodes can be remotely accessed by researchers to run experiments on a catalog of scenarios capturing diverse path loss, shadowing, and fading conditions extracted from real-world wireless environments.

For this work, we consider an urban scenario with 42 \acp{ue} and 7 \acp{bs} whose locations are extracted from the OpenCelliD database~\cite{opencellid} to match that of a real cellular deployment in Rome, Italy. Users are deployed uniformly near the \acp{bs}, which offer connectivity via a $10$~MHz radio channel with a sub-carrier spacing of $15$~KHz. Each \ac{ue} is assigned to a network slice (\ie \ac{embb}, \ac{mmtc}, and \ac{urllc}) and receives traffic according to the slice it belongs to. 
Specifically, we use Colosseum's traffic generator (based on MGEN~\cite{mgen}) to generate two traffic profiles. In the first one (TRF1), \ac{embb} \acp{ue} receive $4$~Mbit/s constant bitrate traffic in \ac{dwl}, while \ac{mmtc} and \ac{urllc} \acp{ue} receive a Poisson traffic in \ac{dwl} with average $44.6$~kbit/s and $89.3$~kbit/s, respectively, as in~\cite{tmc-coloran-wines}. TRF1 is used for generating the training dataset and for live experiments. The second profile (TRF2) features $2$~Mbit/s constant bitrate traffic for \ac{embb}, and $133.9$~kbit/s and $178.6$~kbit/s Poisson traffic for \ac{mmtc} and \ac{urllc}, respectively.

Table~\ref{tab:exp-settings} (in the Appendix~\ref{app-sec:exp-config}) lists the complete set of experiment configurations. All run for 30~minutes. The experiments on the left portion of the table are used in \S~\ref{subsec:motivation-shap} and in \S~\ref{subsec:results-explanations}. The experiments on the right portion of the table are used in \S~\ref{subsec:results-optimization} and run with an additional online training phase, which helps agents adapt to a change in the environment. Overall, we run 28 hours of experiments in Colosseum to gather the data for all the 40 experiments in Table~\ref{tab:exp-settings}. This adds to the more than 150 hours of data collection for the offline dataset used to train the \ac{drl} agents.

\subsection{System Explanations}
\label{subsec:results-explanations}

\begin{figure*}
\captionsetup[subfigure]{captionskip=0pt}
\centering
\hspace*{3em}\includegraphics[scale=.65]{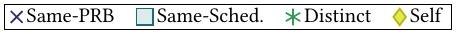}\par
\vspace*{-2.5ex}%
\subfloat[\scriptsize \txbrate and \dlbuff~\label{fig:embb-trf1-txbrate-dlbuffer}]{%
\includegraphics[width=.33\textwidth,keepaspectratio]{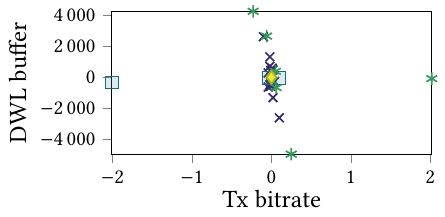}}%
~%
\subfloat[{\scriptsize \txpkts and \dlbuff}~\label{fig:embb-trf1-txpkts-dlbuffer}]{%
\includegraphics[width=.33\textwidth,keepaspectratio]{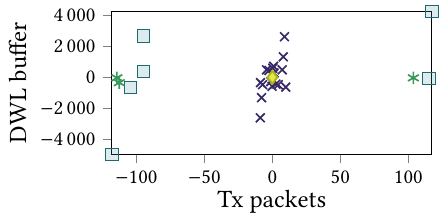}}%
~%
\subfloat[\scriptsize \txbrate and \txpkts~\label{fig:embb-trf1-txbrate-txpkts}]{%
\includegraphics[width=.33\textwidth,keepaspectratio]{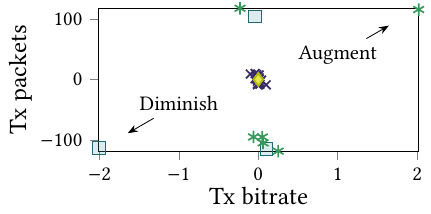}}%
\vspace*{-2ex}%
\caption{Detailed explanations for the \ac{ht} agent's behavior}%
\label{fig:explanation-embb-trf1}%
\vspace*{-4ex}%
\end{figure*}

We distill post-hoc explanations for the \ac{ht} and \ac{ll} agents. Without loss of generality and due to space limitations, Figure~\ref{fig:explanation-embb-trf1} reports the case of TRF1 for the \ac{ht} agent only (the resulting graph for agent \ac{ht} is shown in Appendix~\ref{app-sec:example-graph}). However, the corresponding discussion related to the agent \ac{ll} is given in Appendix~\ref{app-sec:agentll}.
Unlike \ac{shap}, which takes 26~h to generate non-intuitive explanations for experiments with $4+$ users (see \S~\ref{subsec:motivation-shap}), \toolname takes on average only $2.3~s$ to generate, process the graph and synthesize the intuitive explanations relating actions to \acp{kpi} variation ($40695 \times$ faster).

\toolname provides explanations by breaking down at the level of individual \acp{kpi} the effect of transitions between multi-modal actions. 
As actions have two modes, we find in $G$ the following categories of transitions:
\textit{(i)} ``Same-PRB'' is a transition between actions with the same \ac{prb} allocation, \textit{(ii)} ``Same-Sched.'' is a transition between actions with the same scheduling policy, \textit{(iii)} ``Distinct'' is a transition between actions with different \ac{prb} allocation and scheduling policies, and \textit{(iv)} ``Self'' denotes no transition, \ie the same action is repeated in two subsequent steps.

Figure~\ref{fig:explanation-embb-trf1} provides domain experts with the required information to understand how agents operate. Each point of the scatter plots represents a \acp{kpi} variation observed from state $s_{t}$ to $s_{t+1}$ determined by the transition from action $a_{t-1}$ to action $a_t$. Overall, ``Distinct'' transitions produce large variations of \dlbuff while ``Same-PRB'' triggers lower \dlbuff variations with no change in \txbrate (see Figure~\ref{fig:explanation-embb-trf1}\subref{fig:embb-trf1-txbrate-dlbuffer}). Across all agents and settings, ``Self'' and ``Distinct'' transitions are respectively around $5$\% and $50$\% of the total.

To simplify the results in Figure~\ref{fig:explanation-embb-trf1} for non-expert users, we summarize concise explanations that \textit{generate new knowledge} about the reason why the agents use the different categories of actions. In other words, \toolname spots the intertwined relation between \acp{kpi} and how actions determine their change. Figure~\ref{fig:summary-of-exp-embb-trf1} and Table~\ref{tab:summary-of-exp-embb-trf1} show such summary in a more human-friendly form. Specifically, Figure~\ref{fig:summary-of-exp-embb-trf1} is obtained by constructing a \ac{dt} on the data shown in Figure~\ref{fig:explanation-embb-trf1} (\ie the outcome of \toolname) \textit{and not on the agent itself.} As discussed in \S~\ref{subsec:explore-design}, this is necessary as \acp{dt} perform poorly if applied directly to the agent. The new knowledge is generated by tracing the decisions taken at each branch while traversing the tree from the root to the leaves. This process pinpoints the decision-making criteria of the agent for a given category of action transitions. If the same category appears in more than one leaf, then the decision-making process is complex and the same class applies to different \ac{kpi} variations. Surprisingly, the agent uses ``Same-Sched'' to reduce the throughput (\ie \txbrate, see the node on the left branch of the \ac{dt}) and the number of \txpkts (on the right branch of the \ac{dt}, which corresponds to the bottom left point in Figure~\ref{fig:explanation-embb-trf1}\subref{fig:embb-trf1-txbrate-txpkts}. The agent uses ``Same-PRB'' to sustain current throughput by using actions that produce minor variations in the other \acp{kpi} (\txpkts and \dlbuff), and uses ``Self'' when it does not observe any variation \acp{kpi}. Unlike Figure~\ref{fig:explanation-embb-trf1}, Figure~\ref{fig:summary-of-exp-embb-trf1} and Table~\ref{tab:summary-of-exp-embb-trf1} deliver intuitive explanations that are effective to shed light on the agent's behavior.

\medskip
\begin{minipage}{0.53\textwidth}
\centering
\includegraphics[width=.95\textwidth]{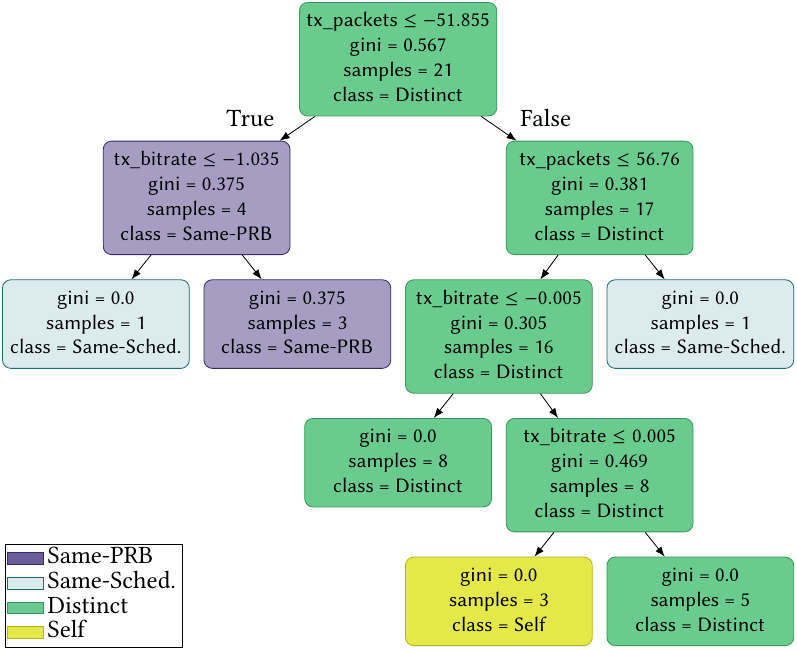}
\captionof{figure}{\ac{dt} on \toolname explanations for the \ac{ht} agent}%
\label{fig:summary-of-exp-embb-trf1}%
\end{minipage}
~%
\begin{minipage}{0.42\textwidth}
\centering
\captionof{table}{\ac{ht} agent: summary of explanations}%
\label{tab:summary-of-exp-embb-trf1}%
\vspace*{-2ex}%
\resizebox{\textwidth}{!}{%
\begin{tabular}{r>{\RaggedRight}p{5.5cm}}%
\toprule
\textsc{Transition} & \textsc{Interpretation}\\
\midrule
\textsc{Same-PRB} & Produces minor changes in \acp{kpi}  \\
\textsc{Same-Sched.} & Diminishes \txbrate, other \acp{kpi} augment/diminish according to the previous state \\
\textsc{Distinct} & Increases \txbrate, other \acp{kpi} augment/diminish according to the previous state \\
\textsc{Self} & No change in \acp{kpi}\\
\bottomrule
\end{tabular}%
}%
\end{minipage}

\subsection{Ad-hoc Adjustments}
\label{subsec:results-optimization}

\begin{figure*}
\captionsetup[subfigure]{captionskip=0pt}
\centering
\vspace*{-2.5ex}%
\subfloat[TRF1~\label{fig:kpi-inc-urllc-trf1}]{%
\includegraphics[width=.49\textwidth,keepaspectratio]{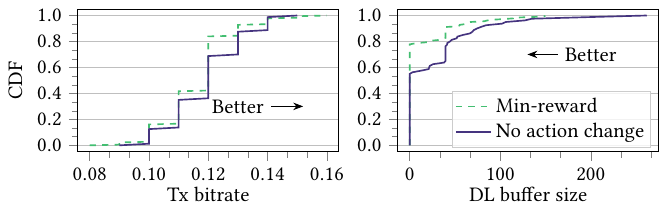}}%
~%
\subfloat[TRF2~\label{fig:kpi-inc-urllc-trf2}]{%
\includegraphics[width=.49\textwidth,keepaspectratio]{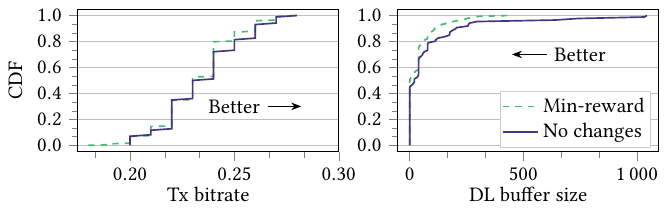}}%
\vspace*{-3ex}%
\caption{\texttt{Min reward} policy effect on \ac{urllc} slice in different traffic scenarios}%
\label{fig:kpi-increase-urllc-trf1_SL2}%
\vspace*{-3ex}%
\end{figure*}

\begin{figure*}
\captionsetup[subfigure]{captionskip=0pt}
\centering
\vspace*{-2.5ex}%
\subfloat[TRF1~\label{fig:kpi-inc-embb-trf1}]{%
\includegraphics[width=.49\textwidth,keepaspectratio]{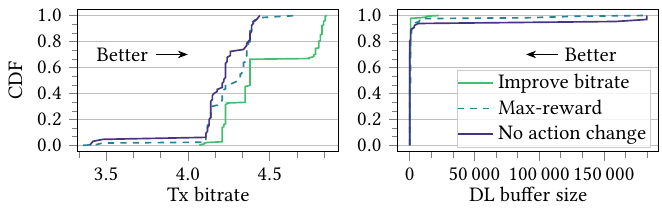}}%
~%
\subfloat[TRF2~\label{fig:kpi-inc-embb-trf2}]{%
\includegraphics[width=.49\textwidth,keepaspectratio]{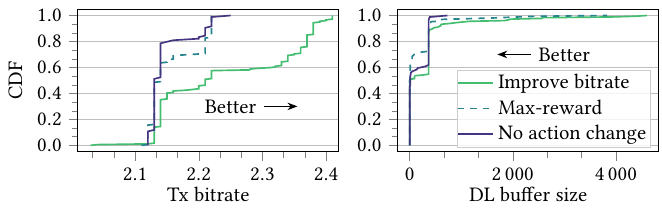}}%
\vspace*{-3ex}%
\caption{\texttt{Max reward} policy and \texttt{improve bitrate} effect on \ac{embb} slice in different traffic scenarios}%
\label{fig:kpi-increase-embb-SL0}%
\vspace*{-3ex}%
\end{figure*}

We now illustrate how explanations can steer the decision-making process toward desired goals defined as intents. To this end, we implement the three policies described in \S~\ref{subsec:impl-ar-strategies} within the \emph{Improve} module of \toolname and benchmark their capability in reducing the \dlbuff for the \ac{ll} agent (``Min-reward'' or \ref{ar:2}, in Figure~\ref{fig:kpi-increase-urllc-trf1_SL2}) and improve \txbrate for the \ac{ht} agent (``Max-reward'' or \ref{ar:1}, and ``Improved bitrate'' or \ref{ar:3} in Figure~\ref{fig:kpi-increase-embb-SL0}). We also show the corresponding counterpart effect on the other \acp{kpi}. We derive the distribution of the \acp{kpi} once $G$ is made available after online training and compare against a baseline without action change. 

``Max-reward'' and ``Improved bitrate'' provide different levels of bitrate improvement (see Figure~\ref{fig:kpi-increase-embb-SL0}): the latter is a more aggressive strategy because it explores the graph looking for first-hop nodes with bitrate attributes directly. On the contrary, ``Max-reward'' changes action solely on the basis of the expected reward that a given first-hop node would provide. The ``Min-reward'' strategy is effective as it reduces significantly the tail of the buffer size occupancy with minor changes in \txbrate, thus allowing for faster transmission of \ac{urllc} traffic (see Figure~\ref{fig:kpi-increase-urllc-trf1_SL2}). Finally, we assert that the intent-based action steering policies do not harm the capabilities of the agent to generalize and adapt to changes in the systems (see~\S~\ref{app-sec:action_steering}).

\section{Discussion on \toolname}
\label{sec:discussion-explora}

\simpletitle{Solving the Challenges} To summarize, the use of attributed graphs solves the challenges presented in \S~\ref{subsec:challenges}. It makes it possible to establish the ``actions-outputs'' link, thereby making the decision-making process observable when it would not be otherwise (\ref{cha:1}); avoids primary cause and effect being undetermined because of memory (\eg it determines whether an action is mainly attributed to the past \ac{prb} allocation or to the change of \acp{kpi}) (\ref{cha:2}); and enables understanding the implications of individual components of multi-modal actions by comparing the distributions of KPIs (attributes) between transitions (edges) of interconnected nodes (actions) with the same or distinct individual multi-modal component (\ref{cha:3}).

\simpletitle{Generalizability} We now comment on how \toolname can generalize within and outside O-RAN networks, e.g., by interfacing with AI/ML algorithms running on the \acp{bs} directly, or on generic network controllers not tied to O-RAN specifications. Further, we will discuss how \toolname can be applied to nodes external to the RAN. The discussion holds as long as the general system architecture is based on the use of autoencoders and \ac{drl} agents, and actions operate on two or more variables as in~\cite{guo-tvt-complexaction,huang-access-complex-action}. Both are examples of agents enforcing multi-modal actions in the \ac{ran}. The multi-agent framework in~\cite{guo-tvt-complexaction} jointly determines \ac{bs} selection and power requirement to maximize user throughput during handovers. The agent in~\cite{huang-access-complex-action} observes the distribution of channel quality indicators per user, amount of data to transmit, and traffic type as state $s$ and takes joint decisions on user scheduling and antenna allocation with multi-modal actions $a$ that hierarchically \textit{(i)} prioritize users, \textit{(ii)} decide the number of antennas per user, and \textit{(iii)} select a precoding algorithm that maximizes spectral efficiency according to the above decisions. If applied to this agent, \toolname would pinpoint the rationale for the three modes of the actions individually and any combination thereof, \eg which precoding algorithm is typically used by the high priority users or how many antennas are typically allocated to regular users.
Aside from O-RAN networks, \toolname can be interfaced with WiFi access points in which \ac{drl}-based closed-control loops are enacted to jointly tune parameters and configurations of the access point (e.g., a multi-modal action handling power control and beam steering). In this case, network metrics and channel measurements (e.g., channel state information with other WiFi nodes) would be first passed through an autoencoder (either running locally or hosted on an external controller) for dimensionality reduction, and then forwarded to a \ac{drl} agent that computes control actions to be enforced at the WiFi access point. Similarly to the previous case, \toolname can be set up to read the input to the autoencoder, and steer the output of the \ac{drl} agent to tailor the network control to specific conditions and environment.

\simpletitle{From Laboratory to Production Environments} \toolname can also be used to speed up the transition of \ac{ai}/\ac{ml} solutions from laboratory to production-like environments. Indeed, xApps can be first developed, pre-trained offline and tested in controlled laboratory environments, thus complying to the O-RAN specifications. Then, they can be transitioned to production-like environments, where they are used to manage complex and large networks. However, the very many production environments where xApps are deployed might not always reflect the laboratory conditions where they were originally pre-trained and tested. Hence, a tool such as \toolname can help steer the actions, and adapt xApps to the novel environments with minimal online re-training, as it usually happens for DRL agents to adapt to new environments~\cite{hamadanian2022reinforcement,tmc-coloran-wines}.

\section{Related Work}
\label{sec:rel-works}
The last years have seen a surge in the uptake of \ac{xai} for \ac{ai}-based networking systems. Seminal works~\cite{zheng2018demystifying,dethise2019cracking} opened the path forward to interpretability in networking contexts. Auric~\cite{auric} and NeuroCuts~\cite{liang2019neural} rely on \acp{dt} that are self-interpretable \ac{ai} models. AT\&T developed Auric to automatically configure \acp{bs} parameters while NeuroCuts performs packet classification with \ac{rl}. Unfortunately, and as mentioned earlier, \acp{dt} are not very effective in complex networking problem such as the \ac{ran}.

\simpletitle{\ac{xai} for Networking} \ac{xai} is at an early stage of conceptualization and adoption in mobile networks~\cite{xai-6g,li-chen-xai-6g, comcom-xai}. The lack of explainability leads to poor AI/ML model design, which might facilitate adversarial attacks~\cite{infocom22zheng,deexp23,infocom23fl}. Metis~\cite{zili-interpret-sigcomm20} interprets diverse \ac{dl}-based networking systems, converting \ac{dnn} policies into interpretable rule-based controllers and highlighting critical components. Metis works well when the relation between \emph{inputs} and \emph{outputs} is explicit, which does not address \ref{cha:1}. The work in \cite{zheng-drl-jsac} uses a teacher-student framework to improve robustness of \ac{drl}-based networks. The teacher infuses a set of white-box logic rules defined by humans into a \ac{drl}-agent, \ie the student. Specifically, the student maximizes the expected cumulative reward while minimizing the distance to the teaching data. Unfortunately, in complex systems like the \ac{ran}, defining white-box logic rules a priori may be prohibitively time-consuming and inefficient. 

The above research efforts still remain preliminary in regard to the design of explainability techniques applicable to complex \ac{ai} models (see \S~\ref{subsec:challenges}) for real networking systems. \toolname fills precisely this gap for the Open \ac{ran} case.

\simpletitle{Robustness} Ensuring model robustness is important and can be done through anomaly detection~\cite{han-deepaid-ccs21} or formal verification~\cite{kazak-fv-drl-netai, marco-fv-infocom}. 
\emph{Shielding} is a safety mechanism that inhibits an agent from taking a potentially risky actions provisioning \ac{drl} agents with an additional layer of robustness. 
Shields may be constructed in post-training by programmatically determining forbidden actions~\cite{alshiekh-shield-drl}, or infer the dynamics of the system and preventing it from reaching hazardous states~\cite{bastani-shiedl-drl}. At training times, shields can restrict the exploration of the state space~\cite{bettina-shield-drl} thereby limiting the agent knowledge right from the start. By contrast, \toolname performs adaptive action-steering.

\section{Conclusions}
\label{sec:conc}

In this paper, we propose \toolname, a new framework to explain the class of resource allocation \ac{drl}-based solutions for cellular networks.
\toolname synthesizes model explanations that facilitate monitoring and troubleshooting for domain experts. We showcase the benefits of \toolname in a typical Open~\ac{ran} scenario and apply it to a set of \ac{drl} agents executing as O-RAN xApps that govern \ac{ran} slicing and scheduling policies. \toolname not only synthesizes clear and concise explanations, but can also improve overall performance by using explanations to proactively identify and substitute actions that would lead to low expected rewards programmatically.

\section*{Acknowledgment}
Dr. Fiandrino visited Northeastern University with the support of a José Castillejo/Fulbright scholarship (CAS21/00500). This work is partially supported by the Spanish Ministry of Science and Innovation through the Juan de la Cierva grant IJC2019-039885-I and grant PID2021-128250NB-I00 (``bRAIN''), by the project RISC-6G, reference TSI-063000-2021-63, granted by the Ministry of Economic Affairs and Digital Transformation and the European Union-NextGenerationEU through the UNICO-5G R\&D Program of the Spanish Recovery, Transformation and Resilience Plan, and by the U.S.\ National Science Foundation under grants CNS-2112471, CNS-1925601, and CNS-2312875.

\bibliographystyle{ACM-Reference-Format}
\bibliography{biblio,oran-biblio}


\begin{thebibliography}{80}


\ifx \showCODEN    \undefined \def \showCODEN     #1{\unskip}     \fi
\ifx \showDOI      \undefined \def \showDOI       #1{#1}\fi
\ifx \showISBNx    \undefined \def \showISBNx     #1{\unskip}     \fi
\ifx \showISBNxiii \undefined \def \showISBNxiii  #1{\unskip}     \fi
\ifx \showISSN     \undefined \def \showISSN      #1{\unskip}     \fi
\ifx \showLCCN     \undefined \def \showLCCN      #1{\unskip}     \fi
\ifx \shownote     \undefined \def \shownote      #1{#1}          \fi
\ifx \showarticletitle \undefined \def \showarticletitle #1{#1}   \fi
\ifx \showURL      \undefined \def \showURL       {\relax}        \fi
\providecommand\bibfield[2]{#2}
\providecommand\bibinfo[2]{#2}
\providecommand\natexlab[1]{#1}
\providecommand\showeprint[2][]{arXiv:#2}

\bibitem[Alshiekh et~al\mbox{.}(2018)]%
        {alshiekh-shield-drl}
\bibfield{author}{\bibinfo{person}{Mohammed Alshiekh},
  \bibinfo{person}{Roderick Bloem}, \bibinfo{person}{Rüdiger Ehlers},
  \bibinfo{person}{Bettina Könighofer}, \bibinfo{person}{Scott Niekum}, {and}
  \bibinfo{person}{Ufuk Topcu}.} \bibinfo{year}{2018}\natexlab{}.
\newblock \showarticletitle{Safe Reinforcement Learning via Shielding}.
\newblock \bibinfo{journal}{\emph{Proc.~of AAAI Conference on Artificial
  Intelligence}} \bibinfo{volume}{32}, \bibinfo{number}{1}
  (\bibinfo{date}{Apr.} \bibinfo{year}{2018}).
\newblock
\urldef\tempurl%
\url{https://doi.org/10.1609/aaai.v32i1.11797}
\showDOI{\tempurl}


\bibitem[Ayala-Romero et~al\mbox{.}(2019)]%
        {vrain-andres-mobicom19}
\bibfield{author}{\bibinfo{person}{Jose~A. Ayala-Romero},
  \bibinfo{person}{Andres Garcia-Saavedra}, \bibinfo{person}{Marco Gramaglia},
  \bibinfo{person}{Xavier Costa-Perez}, \bibinfo{person}{Albert Banchs}, {and}
  \bibinfo{person}{Juan~J. Alcaraz}.} \bibinfo{year}{2019}\natexlab{}.
\newblock \showarticletitle{{VrAIn}: A Deep Learning Approach Tailoring
  Computing and Radio Resources in Virtualized {RANs}}. In
  \bibinfo{booktitle}{\emph{Proc.~of~ACM MobiCom}}.
\newblock
\showISBNx{9781450361699}
\urldef\tempurl%
\url{https://doi.org/10.1145/3300061.3345431}
\showDOI{\tempurl}


\bibitem[Baldesi et~al\mbox{.}(2022)]%
        {infocom22-luca-wines}
\bibfield{author}{\bibinfo{person}{Luca Baldesi}, \bibinfo{person}{Francesco
  Restuccia}, {and} \bibinfo{person}{Tommaso Melodia}.}
  \bibinfo{year}{2022}\natexlab{}.
\newblock \showarticletitle{{ChARM}: {NextG} Spectrum Sharing Through
  Data-Driven Real-Time O-RAN Dynamic Control}. In
  \bibinfo{booktitle}{\emph{Proc.~of~IEEE INFOCOM}}. \bibinfo{pages}{240--249}.
\newblock
\urldef\tempurl%
\url{https://doi.org/10.1109/INFOCOM48880.2022.9796985}
\showDOI{\tempurl}


\bibitem[Bastani(2021)]%
        {bastani-shiedl-drl}
\bibfield{author}{\bibinfo{person}{Osbert Bastani}.}
  \bibinfo{year}{2021}\natexlab{}.
\newblock \showarticletitle{Safe Reinforcement Learning with Nonlinear Dynamics
  via Model Predictive Shielding}. In \bibinfo{booktitle}{\emph{Proc.~of
  American Control Conference}}. \bibinfo{pages}{3488--3494}.
\newblock
\urldef\tempurl%
\url{https://doi.org/10.23919/ACC50511.2021.9483182}
\showDOI{\tempurl}


\bibitem[Bertizzolo et~al\mbox{.}(2021)]%
        {tmc-lorenzo-wines}
\bibfield{author}{\bibinfo{person}{Lorenzo Bertizzolo},
  \bibinfo{person}{Tuyen~Xuan Tran}, \bibinfo{person}{John Buczek},
  \bibinfo{person}{Bharath Balasubramanian}, \bibinfo{person}{Rittwik Jana},
  \bibinfo{person}{Yu Zhou}, {and} \bibinfo{person}{Tommaso Melodia}.}
  \bibinfo{year}{2021}\natexlab{}.
\newblock \showarticletitle{Streaming from the Air: Enabling Drone-sourced
  Video Streaming Applications on {5G Open-RAN} Architectures}.
\newblock \bibinfo{journal}{\emph{IEEE Transactions on Mobile Computing}}
  (\bibinfo{year}{2021}), \bibinfo{pages}{1--1}.
\newblock
\urldef\tempurl%
\url{https://doi.org/10.1109/TMC.2021.3129094}
\showDOI{\tempurl}


\bibitem[Bonati et~al\mbox{.}(2022)]%
        {openrangym}
\bibfield{author}{\bibinfo{person}{Leonardo Bonati}, \bibinfo{person}{Michele
  Polese}, \bibinfo{person}{Salvatore D’Oro}, \bibinfo{person}{Stefano
  Basagni}, {and} \bibinfo{person}{Tommaso Melodia}.}
  \bibinfo{year}{2022}\natexlab{}.
\newblock \showarticletitle{{OpenRAN Gym}: An Open Toolbox for Data Collection
  and Experimentation with {AI} in {O-RAN}}. In
  \bibinfo{booktitle}{\emph{Proc.~of~IEEE WCNC}}. \bibinfo{pages}{518--523}.
\newblock
\urldef\tempurl%
\url{https://doi.org/10.1109/WCNC51071.2022.9771908}
\showDOI{\tempurl}


\bibitem[Broniatowski et~al\mbox{.}(2021)]%
        {broniatowski2021psychological}
\bibfield{author}{\bibinfo{person}{David~A Broniatowski} {et~al\mbox{.}}}
  \bibinfo{year}{2021}\natexlab{}.
\newblock \showarticletitle{Psychological foundations of explainability and
  interpretability in artificial intelligence}.
\newblock \bibinfo{journal}{\emph{NIST, Tech. Rep}} (\bibinfo{year}{2021}).
\newblock


\bibitem[Chen et~al\mbox{.}(2023)]%
        {chen-shield-example-drl}
\bibfield{author}{\bibinfo{person}{Pengcheng Chen}, \bibinfo{person}{Shichao
  Liu}, \bibinfo{person}{Xiaozhe Wang}, {and} \bibinfo{person}{Innocent
  Kamwa}.} \bibinfo{year}{2023}\natexlab{}.
\newblock \showarticletitle{Physics-Shielded Multi-Agent Deep Reinforcement
  Learning for Safe Active Voltage Control With Photovoltaic/Battery Energy
  Storage Systems}.
\newblock \bibinfo{journal}{\emph{IEEE Transactions on Smart Grid}}
  \bibinfo{volume}{14}, \bibinfo{number}{4} (\bibinfo{year}{2023}),
  \bibinfo{pages}{2656--2667}.
\newblock
\urldef\tempurl%
\url{https://doi.org/10.1109/TSG.2022.3228636}
\showDOI{\tempurl}


\bibitem[Chen and Guestrin(2016)]%
        {chen2016xgboost}
\bibfield{author}{\bibinfo{person}{Tianqi Chen} {and} \bibinfo{person}{Carlos
  Guestrin}.} \bibinfo{year}{2016}\natexlab{}.
\newblock \showarticletitle{{XGboost}: A scalable tree boosting system}. In
  \bibinfo{booktitle}{\emph{Proc.~of ACM SIGKDD}}. \bibinfo{pages}{785--794}.
\newblock


\bibitem[Dazeley et~al\mbox{.}(2021)]%
        {dazeley-xrl}
\bibfield{author}{\bibinfo{person}{Richard Dazeley}, \bibinfo{person}{Peter
  Vamplew}, {and} \bibinfo{person}{Francisco Cruz}.}
  \bibinfo{year}{2021}\natexlab{}.
\newblock \showarticletitle{Explainable reinforcement learning for broad-{XAI}:
  a conceptual framework and survey}.
\newblock \bibinfo{journal}{\emph{arXiv preprint arXiv:2108.09003}}
  (\bibinfo{year}{2021}).
\newblock


\bibitem[Dethise et~al\mbox{.}(2019)]%
        {dethise2019cracking}
\bibfield{author}{\bibinfo{person}{Arnaud Dethise}, \bibinfo{person}{Marco
  Canini}, {and} \bibinfo{person}{Srikanth Kandula}.}
  \bibinfo{year}{2019}\natexlab{}.
\newblock \showarticletitle{Cracking open the black box: What observations can
  tell us about reinforcement learning agents}. In
  \bibinfo{booktitle}{\emph{Proc.~of ACM NetAI}}. \bibinfo{pages}{29--36}.
\newblock


\bibitem[Dethise et~al\mbox{.}(2021)]%
        {marco-fv-infocom}
\bibfield{author}{\bibinfo{person}{Arnaud Dethise}, \bibinfo{person}{Marco
  Canini}, {and} \bibinfo{person}{Nina Narodytska}.}
  \bibinfo{year}{2021}\natexlab{}.
\newblock \showarticletitle{Analyzing Learning-Based Networked Systems with
  Formal Verification}. In \bibinfo{booktitle}{\emph{Proc.~of~IEEE~INFOCOM}}.
  \bibinfo{pages}{1--10}.
\newblock
\urldef\tempurl%
\url{https://doi.org/10.1109/INFOCOM42981.2021.9488898}
\showDOI{\tempurl}


\bibitem[Dryjański et~al\mbox{.}(2021)]%
        {mdpisensors-trafficsteeringoran}
\bibfield{author}{\bibinfo{person}{Marcin Dryjański}, \bibinfo{person}{Łukasz
  Kułacz}, {and} \bibinfo{person}{Adrian Kliks}.}
  \bibinfo{year}{2021}\natexlab{}.
\newblock \showarticletitle{Toward Modular and Flexible Open {RAN}
  Implementations in {6G} Networks: Traffic Steering Use Case and {O-RAN
  xApps}}.
\newblock \bibinfo{journal}{\emph{Sensors}} \bibinfo{volume}{21},
  \bibinfo{number}{24} (\bibinfo{year}{2021}).
\newblock
\showISSN{1424-8220}
\urldef\tempurl%
\url{https://doi.org/10.3390/s21248173}
\showDOI{\tempurl}


\bibitem[Du et~al\mbox{.}(2019)]%
        {du2019techniques}
\bibfield{author}{\bibinfo{person}{Mengnan Du}, \bibinfo{person}{Ninghao Liu},
  {and} \bibinfo{person}{Xia Hu}.} \bibinfo{year}{2019}\natexlab{}.
\newblock \showarticletitle{Techniques for interpretable machine learning}.
\newblock \bibinfo{journal}{\emph{Commun. ACM}} \bibinfo{volume}{63},
  \bibinfo{number}{1} (\bibinfo{year}{2019}), \bibinfo{pages}{68--77}.
\newblock


\bibitem[Fei et~al\mbox{.}(2023)]%
        {infocom23fl}
\bibfield{author}{\bibinfo{person}{Wang Fei}, \bibinfo{person}{Hugh Ethan},
  {and} \bibinfo{person}{Li Baochun}.} \bibinfo{year}{2023}\natexlab{}.
\newblock \showarticletitle{More than Enough is Too Much: Adaptive Defenses
  against Gradient Leakage in Production Federated Learning}. In
  \bibinfo{booktitle}{\emph{Proc.~of IEEE INFOCOM}}. \bibinfo{pages}{1--10}.
\newblock
\urldef\tempurl%
\url{https://doi.org/10.1109/INFOCOM53939.2023.10228919}
\showDOI{\tempurl}


\bibitem[Fiandrino et~al\mbox{.}(2022)]%
        {comcom-xai}
\bibfield{author}{\bibinfo{person}{Claudio Fiandrino}, \bibinfo{person}{Giulia
  Attanasio}, \bibinfo{person}{Marco Fiore}, {and} \bibinfo{person}{Joerg
  Widmer}.} \bibinfo{year}{2022}\natexlab{}.
\newblock \showarticletitle{Toward native explainable and robust {AI} in {6G}
  networks: Current state, challenges and road ahead}.
\newblock \bibinfo{journal}{\emph{Computer Communications}}
  \bibinfo{volume}{193} (\bibinfo{year}{2022}), \bibinfo{pages}{47--52}.
\newblock
\showISSN{0140-3664}
\urldef\tempurl%
\url{https://doi.org/10.1016/j.comcom.2022.06.036}
\showDOI{\tempurl}


\bibitem[Garcia-Saavedra and Costa-Pérez(2021)]%
        {csm21-andres}
\bibfield{author}{\bibinfo{person}{Andres Garcia-Saavedra} {and}
  \bibinfo{person}{Xavier Costa-Pérez}.} \bibinfo{year}{2021}\natexlab{}.
\newblock \showarticletitle{{O-RAN}: Disrupting the Virtualized {RAN}
  Ecosystem}.
\newblock \bibinfo{journal}{\emph{IEEE Communications Standards Magazine}}
  \bibinfo{volume}{5}, \bibinfo{number}{4} (\bibinfo{year}{2021}),
  \bibinfo{pages}{96--103}.
\newblock
\urldef\tempurl%
\url{https://doi.org/10.1109/MCOMSTD.101.2000014}
\showDOI{\tempurl}


\bibitem[Ge et~al\mbox{.}(2023)]%
        {ge-chroma-mobicom23}
\bibfield{author}{\bibinfo{person}{Changhan Ge}, \bibinfo{person}{Zihui Ge},
  \bibinfo{person}{Xuan Liu}, \bibinfo{person}{Ajay Mahimkar},
  \bibinfo{person}{Yusef Shaqalle}, \bibinfo{person}{Yu Xiang}, {and}
  \bibinfo{person}{Shomik Pathak}.} \bibinfo{year}{2023}\natexlab{}.
\newblock \showarticletitle{Chroma: Learning and Using Network Contexts to
  Reinforce Performance Improving Configurations}. In
  \bibinfo{booktitle}{\emph{Proc.~of~ACM MobiCom}}.
\newblock
\showISBNx{9781450399906}
\urldef\tempurl%
\url{https://doi.org/10.1145/3570361.3613256}
\showDOI{\tempurl}


\bibitem[Guidotti et~al\mbox{.}(2018)]%
        {guidotti2018survey}
\bibfield{author}{\bibinfo{person}{Riccardo Guidotti}, \bibinfo{person}{Anna
  Monreale}, \bibinfo{person}{Salvatore Ruggieri}, \bibinfo{person}{Franco
  Turini}, \bibinfo{person}{Fosca Giannotti}, {and} \bibinfo{person}{Dino
  Pedreschi}.} \bibinfo{year}{2018}\natexlab{}.
\newblock \showarticletitle{A survey of methods for explaining black box
  models}.
\newblock \bibinfo{journal}{\emph{ACM computing surveys (CSUR)}}
  \bibinfo{volume}{51}, \bibinfo{number}{5} (\bibinfo{year}{2018}),
  \bibinfo{pages}{1--42}.
\newblock


\bibitem[Gunning and Aha(2019)]%
        {gunning2019darpa}
\bibfield{author}{\bibinfo{person}{David Gunning} {and} \bibinfo{person}{David
  Aha}.} \bibinfo{year}{2019}\natexlab{}.
\newblock \showarticletitle{DARPA's explainable artificial intelligence (XAI)
  program}.
\newblock \bibinfo{journal}{\emph{AI magazine}} \bibinfo{volume}{40},
  \bibinfo{number}{2} (\bibinfo{year}{2019}), \bibinfo{pages}{44--58}.
\newblock


\bibitem[Guo et~al\mbox{.}(2020)]%
        {guo-tvt-complexaction}
\bibfield{author}{\bibinfo{person}{Delin Guo}, \bibinfo{person}{Lan Tang},
  \bibinfo{person}{Xinggan Zhang}, {and} \bibinfo{person}{Ying-Chang Liang}.}
  \bibinfo{year}{2020}\natexlab{}.
\newblock \showarticletitle{Joint Optimization of Handover Control and Power
  Allocation Based on Multi-Agent Deep Reinforcement Learning}.
\newblock \bibinfo{journal}{\emph{IEEE Transactions on Vehicular Technology}}
  \bibinfo{volume}{69}, \bibinfo{number}{11} (\bibinfo{year}{2020}),
  \bibinfo{pages}{13124--13138}.
\newblock
\urldef\tempurl%
\url{https://doi.org/10.1109/TVT.2020.3020400}
\showDOI{\tempurl}


\bibitem[Guo(2020)]%
        {xai-6g}
\bibfield{author}{\bibinfo{person}{Weisi Guo}.}
  \bibinfo{year}{2020}\natexlab{}.
\newblock \showarticletitle{Explainable Artificial Intelligence for {6G}:
  Improving Trust between Human and Machine}.
\newblock \bibinfo{journal}{\emph{IEEE Communications Magazine}}
  \bibinfo{volume}{58}, \bibinfo{number}{6} (\bibinfo{year}{2020}),
  \bibinfo{pages}{39--45}.
\newblock
\urldef\tempurl%
\url{https://doi.org/10.1109/MCOM.001.2000050}
\showDOI{\tempurl}


\bibitem[Hamadanian et~al\mbox{.}(2022)]%
        {hamadanian2022reinforcement}
\bibfield{author}{\bibinfo{person}{Pouya Hamadanian}, \bibinfo{person}{Malte
  Schwarzkopf}, \bibinfo{person}{Siddartha Sen}, {and}
  \bibinfo{person}{Mohammad Alizadeh}.} \bibinfo{year}{2022}\natexlab{}.
\newblock \showarticletitle{How Reinforcement Learning Systems Fail and What to
  do About It}. In \bibinfo{booktitle}{\emph{Proc.~of~EuroMLSys}}.
\newblock


\bibitem[Han et~al\mbox{.}(2021)]%
        {han-deepaid-ccs21}
\bibfield{author}{\bibinfo{person}{Dongqi Han}, \bibinfo{person}{Zhiliang
  Wang}, \bibinfo{person}{Wenqi Chen}, \bibinfo{person}{Ying Zhong},
  \bibinfo{person}{Su Wang}, \bibinfo{person}{Han Zhang},
  \bibinfo{person}{Jiahai Yang}, \bibinfo{person}{Xingang Shi}, {and}
  \bibinfo{person}{Xia Yin}.} \bibinfo{year}{2021}\natexlab{}.
\newblock \showarticletitle{{DeepAID}: Interpreting and Improving Deep
  Learning-Based Anomaly Detection in Security Applications}. In
  \bibinfo{booktitle}{\emph{Proc.~of ACM SIGSAC CCS}}.
  \bibinfo{pages}{3197–3217}.
\newblock
\showISBNx{9781450384544}
\urldef\tempurl%
\url{https://doi.org/10.1145/3460120.3484589}
\showDOI{\tempurl}


\bibitem[He et~al\mbox{.}(2021)]%
        {HE2021107052}
\bibfield{author}{\bibinfo{person}{Lei He}, \bibinfo{person}{Nabil Aouf}, {and}
  \bibinfo{person}{Bifeng Song}.} \bibinfo{year}{2021}\natexlab{}.
\newblock \showarticletitle{Explainable Deep Reinforcement Learning for {UAV}
  autonomous path planning}.
\newblock \bibinfo{journal}{\emph{Aerospace Science and Technology}}
  \bibinfo{volume}{118} (\bibinfo{year}{2021}), \bibinfo{pages}{107052}.
\newblock
\showISSN{1270-9638}
\urldef\tempurl%
\url{https://doi.org/10.1016/j.ast.2021.107052}
\showDOI{\tempurl}


\bibitem[Heuillet et~al\mbox{.}(2022)]%
        {alexandre-shap-coll-multiagent-drl-mag}
\bibfield{author}{\bibinfo{person}{Alexandre Heuillet}, \bibinfo{person}{Fabien
  Couthouis}, {and} \bibinfo{person}{Natalia D\'{\i}az-Rodr\'{\i}guez}.}
  \bibinfo{year}{2022}\natexlab{}.
\newblock \showarticletitle{{Collective EXplainable AI}: Explaining Cooperative
  Strategies and Agent Contribution in Multiagent Reinforcement Learning With
  Shapley Values}.
\newblock \bibinfo{journal}{\emph{IEEE Comp. Intell. Mag.}}
  \bibinfo{volume}{17}, \bibinfo{number}{1} (\bibinfo{date}{Feb}
  \bibinfo{year}{2022}), \bibinfo{pages}{59–71}.
\newblock
\showISSN{1556-603X}
\urldef\tempurl%
\url{https://doi.org/10.1109/MCI.2021.3129959}
\showDOI{\tempurl}


\bibitem[Ho and Nguyen(2022)]%
        {ho2022joint}
\bibfield{author}{\bibinfo{person}{Tai~Manh Ho} {and} \bibinfo{person}{Kim-Khoa
  Nguyen}.} \bibinfo{year}{2022}\natexlab{}.
\newblock \showarticletitle{{Joint Server Selection, Cooperative Offloading and
  Handover in Multi-Access Edge Computing Wireless Network: A Deep
  Reinforcement Learning Approach}}.
\newblock \bibinfo{journal}{\emph{IEEE Transactions on Mobile Computing}}
  \bibinfo{volume}{21}, \bibinfo{number}{7} (\bibinfo{date}{July}
  \bibinfo{year}{2022}), \bibinfo{pages}{2421--2435}.
\newblock
\showISSN{1558-0660}


\bibitem[Huang et~al\mbox{.}(2023)]%
        {huang-access-complex-action}
\bibfield{author}{\bibinfo{person}{Chih-Wei Huang}, \bibinfo{person}{Ibrahim
  Althamary}, \bibinfo{person}{Yen-Cheng Chou}, \bibinfo{person}{Hong-Yunn
  Chen}, {and} \bibinfo{person}{Cheng-Fu Chou}.}
  \bibinfo{year}{2023}\natexlab{}.
\newblock \showarticletitle{A {DRL}-Based Automated Algorithm Selection
  Framework for Cross-Layer {QoS}-Aware Scheduling and Antenna Allocation in
  Massive MIMO Systems}.
\newblock \bibinfo{journal}{\emph{IEEE Access}}  \bibinfo{volume}{11}
  (\bibinfo{year}{2023}), \bibinfo{pages}{13243--13256}.
\newblock
\urldef\tempurl%
\url{https://doi.org/10.1109/ACCESS.2023.3243068}
\showDOI{\tempurl}


\bibitem[{Intel}(2022)]%
        {intel-drl}
\bibfield{author}{\bibinfo{person}{{Intel}}.} \bibinfo{year}{2022}\natexlab{}.
\newblock \bibinfo{title}{{Intelligent Connection Management for Automated
  Handover Reference Implementation}}.
\newblock
  \bibinfo{howpublished}{\url{https://www.intel.com/content/www/us/en/developer/articles/reference-implementation/intelligent-connection-management.html}}.
\newblock


\bibitem[Iturria-Rivera et~al\mbox{.}(2022)]%
        {iturria2022multi}
\bibfield{author}{\bibinfo{person}{Pedro~Enrique Iturria-Rivera},
  \bibinfo{person}{Han Zhang}, \bibinfo{person}{Hao Zhou},
  \bibinfo{person}{Shahram Mollahasani}, {and} \bibinfo{person}{Melike
  Erol-Kantarci}.} \bibinfo{year}{2022}\natexlab{}.
\newblock \showarticletitle{Multi-agent team learning in virtualized open radio
  access networks (O-RAN)}.
\newblock \bibinfo{journal}{\emph{Sensors}} \bibinfo{volume}{22},
  \bibinfo{number}{14} (\bibinfo{year}{2022}), \bibinfo{pages}{5375}.
\newblock


\bibitem[Jacobs et~al\mbox{.}(2022)]%
        {trustee}
\bibfield{author}{\bibinfo{person}{Arthur~S. Jacobs}, \bibinfo{person}{Roman
  Beltiukov}, \bibinfo{person}{Walter Willinger}, \bibinfo{person}{Ronaldo~A.
  Ferreira}, \bibinfo{person}{Arpit Gupta}, {and} \bibinfo{person}{Lisandro~Z.
  Granville}.} \bibinfo{year}{2022}\natexlab{}.
\newblock \showarticletitle{{AI/ML} for Network Security: The Emperor Has No
  Clothes}. In \bibinfo{booktitle}{\emph{Proc.~of~ACM SIGSAC CCS}}.
  \bibinfo{pages}{1537–1551}.
\newblock
\showISBNx{9781450394505}
\urldef\tempurl%
\url{https://doi.org/10.1145/3548606.3560609}
\showDOI{\tempurl}


\bibitem[Johnson et~al\mbox{.}(2022)]%
        {wintech-nextran}
\bibfield{author}{\bibinfo{person}{David Johnson}, \bibinfo{person}{Dustin
  Maas}, {and} \bibinfo{person}{Jacobus Van Der~Merwe}.}
  \bibinfo{year}{2022}\natexlab{}.
\newblock \showarticletitle{{NexRAN}: Closed-Loop {RAN} Slicing in {POWDER} - A
  Top-to-Bottom Open-Source {Open-RAN} Use Case}. In
  \bibinfo{booktitle}{\emph{Proc.~of ACM WiNTECH}}. \bibinfo{pages}{17–23}.
\newblock
\showISBNx{9781450387033}
\urldef\tempurl%
\url{https://doi.org/10.1145/3477086.3480842}
\showDOI{\tempurl}


\bibitem[Kazak et~al\mbox{.}(2019)]%
        {kazak-fv-drl-netai}
\bibfield{author}{\bibinfo{person}{Yafim Kazak}, \bibinfo{person}{Clark
  Barrett}, \bibinfo{person}{Guy Katz}, {and} \bibinfo{person}{Michael
  Schapira}.} \bibinfo{year}{2019}\natexlab{}.
\newblock \showarticletitle{Verifying {Deep-RL}-Driven Systems}. In
  \bibinfo{booktitle}{\emph{Proc.~of ACM NetAI}}. \bibinfo{pages}{83–89}.
\newblock
\showISBNx{9781450368728}
\urldef\tempurl%
\url{https://doi.org/10.1145/3341216.3342218}
\showDOI{\tempurl}


\bibitem[K{\"o}nighofer et~al\mbox{.}(2020)]%
        {bettina-shield-drl}
\bibfield{author}{\bibinfo{person}{Bettina K{\"o}nighofer},
  \bibinfo{person}{Florian Lorber}, \bibinfo{person}{Nils Jansen}, {and}
  \bibinfo{person}{Roderick Bloem}.} \bibinfo{year}{2020}\natexlab{}.
\newblock \showarticletitle{Shield Synthesis for Reinforcement Learning}. In
  \bibinfo{booktitle}{\emph{Leveraging Applications of Formal Methods,
  Verification and Validation: Verification Principles}},
  \bibfield{editor}{\bibinfo{person}{Tiziana Margaria} {and}
  \bibinfo{person}{Bernhard Steffen}} (Eds.). \bibinfo{publisher}{Springer
  International Publishing}, \bibinfo{pages}{290--306}.
\newblock
\showISBNx{978-3-030-61362-4}


\bibitem[Korobov and Lopuhin(2021)]%
        {eli5}
\bibfield{author}{\bibinfo{person}{M. Korobov} {and} \bibinfo{person}{K.
  Lopuhin}.} \bibinfo{year}{2021}\natexlab{}.
\newblock \bibinfo{title}{{ELI5} is a Python library - v. 0.11}.
\newblock
\newblock
\newblock
\shownote{Available at (accessed 26/10/2021):
  \url{https://eli5.readthedocs.io/en/latest/}}.


\bibitem[Krajna et~al\mbox{.}(2022)]%
        {krajna-xrl}
\bibfield{author}{\bibinfo{person}{Agneza Krajna}, \bibinfo{person}{Mario
  Brcic}, \bibinfo{person}{Tomislav Lipic}, {and} \bibinfo{person}{Juraj
  Doncevic}.} \bibinfo{year}{2022}\natexlab{}.
\newblock \showarticletitle{Explainability in reinforcement learning:
  perspective and position}.
\newblock \bibinfo{journal}{\emph{arXiv preprint arXiv:2203.11547}}
  (\bibinfo{year}{2022}).
\newblock


\bibitem[Laboratory et~al\mbox{.}(2013)]%
        {lawrence2013attributed}
\bibfield{author}{\bibinfo{person}{Lawrence Berkeley~National Laboratory},
  \bibinfo{person}{United States.~Department of~Energy. Office~of Scientific},
  {and} \bibinfo{person}{Technical Information}.}
  \bibinfo{year}{2013}\natexlab{}.
\newblock \bibinfo{booktitle}{\emph{Attributed Graph Models: Modeling Network
  Structure with Correlated Attributes}}.
\newblock \bibinfo{publisher}{United States. Department of Energy.}
\newblock


\bibitem[Lange and Riedmiller(2010)]%
        {ac-for-drl}
\bibfield{author}{\bibinfo{person}{Sascha Lange} {and} \bibinfo{person}{Martin
  Riedmiller}.} \bibinfo{year}{2010}\natexlab{}.
\newblock \showarticletitle{Deep auto-encoder neural networks in reinforcement
  learning}. In \bibinfo{booktitle}{\emph{International Joint Conference on
  Neural Networks (IJCNN)}}. \bibinfo{pages}{1--8}.
\newblock
\urldef\tempurl%
\url{https://doi.org/10.1109/IJCNN.2010.5596468}
\showDOI{\tempurl}


\bibitem[Leivadeas and Falkner(2023)]%
        {aris-ibn-comst}
\bibfield{author}{\bibinfo{person}{Aris Leivadeas} {and}
  \bibinfo{person}{Matthias Falkner}.} \bibinfo{year}{2023}\natexlab{}.
\newblock \showarticletitle{A Survey on Intent-Based Networking}.
\newblock \bibinfo{journal}{\emph{IEEE Communications Surveys \& Tutorials}}
  \bibinfo{volume}{25}, \bibinfo{number}{1} (\bibinfo{year}{2023}),
  \bibinfo{pages}{625--655}.
\newblock
\urldef\tempurl%
\url{https://doi.org/10.1109/COMST.2022.3215919}
\showDOI{\tempurl}


\bibitem[Li et~al\mbox{.}(2020)]%
        {li-chen-xai-6g}
\bibfield{author}{\bibinfo{person}{Chen Li}, \bibinfo{person}{Weisi Guo},
  \bibinfo{person}{Schyler~Chengyao Sun}, \bibinfo{person}{Saba Al-Rubaye},
  {and} \bibinfo{person}{Antonios Tsourdos}.} \bibinfo{year}{2020}\natexlab{}.
\newblock \showarticletitle{Trustworthy Deep Learning in {6G}-Enabled Mass
  Autonomy: From Concept to Quality-of-Trust Key Performance Indicators}.
\newblock \bibinfo{journal}{\emph{IEEE Vehicular Technology Magazine}}
  \bibinfo{volume}{15}, \bibinfo{number}{4} (\bibinfo{year}{2020}),
  \bibinfo{pages}{112--121}.
\newblock
\urldef\tempurl%
\url{https://doi.org/10.1109/MVT.2020.3017181}
\showDOI{\tempurl}


\bibitem[Li et~al\mbox{.}(2021a)]%
        {li-shap-multiagent-drl-kdd}
\bibfield{author}{\bibinfo{person}{Jiahui Li}, \bibinfo{person}{Kun Kuang},
  \bibinfo{person}{Baoxiang Wang}, \bibinfo{person}{Furui Liu},
  \bibinfo{person}{Long Chen}, \bibinfo{person}{Fei Wu}, {and}
  \bibinfo{person}{Jun Xiao}.} \bibinfo{year}{2021}\natexlab{a}.
\newblock \showarticletitle{Shapley Counterfactual Credits for Multi-Agent
  Reinforcement Learning}. In \bibinfo{booktitle}{\emph{Proc.~of~ACM~SIGKDD}}.
  \bibinfo{pages}{934–942}.
\newblock
\showISBNx{9781450383325}
\urldef\tempurl%
\url{https://doi.org/10.1145/3447548.3467420}
\showDOI{\tempurl}


\bibitem[Li et~al\mbox{.}(2021b)]%
        {li2021rlops}
\bibfield{author}{\bibinfo{person}{Peizheng Li}, \bibinfo{person}{Jonathan
  Thomas}, \bibinfo{person}{Xiaoyang Wang}, \bibinfo{person}{Ahmed Khalil},
  \bibinfo{person}{Abdelrahim Ahmad}, \bibinfo{person}{Rui Inacio},
  \bibinfo{person}{Shipra Kapoor}, \bibinfo{person}{Arjun Parekh},
  \bibinfo{person}{Angela Doufexi}, \bibinfo{person}{Arman Shojaeifard},
  {et~al\mbox{.}}} \bibinfo{year}{2021}\natexlab{b}.
\newblock \showarticletitle{{RLOps}: Development Life-cycle of Reinforcement
  Learning Aided {Open RAN}}.
\newblock \bibinfo{journal}{\emph{arXiv preprint}} (\bibinfo{year}{2021}).
\newblock


\bibitem[Li et~al\mbox{.}(2023)]%
        {li2023tapfinger}
\bibfield{author}{\bibinfo{person}{Yihong Li}, \bibinfo{person}{Tianyu Zeng},
  \bibinfo{person}{Xiaoxi Zhang}, \bibinfo{person}{Jingpu Duan}, {and}
  \bibinfo{person}{Chuan Wu}.} \bibinfo{year}{2023}\natexlab{}.
\newblock \showarticletitle{TapFinger: Task Placement and Fine-Grained Resource
  Allocation for Edge Machine Learning}. In \bibinfo{booktitle}{\emph{Proc.~of
  IEEE INFOCOM}}. \bibinfo{pages}{1--10}.
\newblock
\urldef\tempurl%
\url{https://doi.org/10.1109/INFOCOM53939.2023.10229031}
\showDOI{\tempurl}


\bibitem[Liang et~al\mbox{.}(2019)]%
        {liang2019neural}
\bibfield{author}{\bibinfo{person}{Eric Liang}, \bibinfo{person}{Hang Zhu},
  \bibinfo{person}{Xin Jin}, {and} \bibinfo{person}{Ion Stoica}.}
  \bibinfo{year}{2019}\natexlab{}.
\newblock \showarticletitle{Neural packet classification}.
\newblock In \bibinfo{booktitle}{\emph{Proc.~of~ACM SIGCOMM}}.
  \bibinfo{pages}{256--269}.
\newblock


\bibitem[Lundberg et~al\mbox{.}(2020)]%
        {lundberg2020local}
\bibfield{author}{\bibinfo{person}{Scott~M Lundberg}, \bibinfo{person}{Gabriel
  Erion}, \bibinfo{person}{Hugh Chen}, \bibinfo{person}{Alex DeGrave},
  \bibinfo{person}{Jordan~M Prutkin}, \bibinfo{person}{Bala Nair},
  \bibinfo{person}{Ronit Katz}, \bibinfo{person}{Jonathan Himmelfarb},
  \bibinfo{person}{Nisha Bansal}, {and} \bibinfo{person}{Su-In Lee}.}
  \bibinfo{year}{2020}\natexlab{}.
\newblock \showarticletitle{From local explanations to global understanding
  with explainable AI for trees}.
\newblock \bibinfo{journal}{\emph{Nature machine intelligence}}
  \bibinfo{volume}{2}, \bibinfo{number}{1} (\bibinfo{year}{2020}),
  \bibinfo{pages}{56--67}.
\newblock


\bibitem[Lundberg and Lee(2017)]%
        {shap-lundberg2017unified}
\bibfield{author}{\bibinfo{person}{Scott~M Lundberg} {and}
  \bibinfo{person}{Su-In Lee}.} \bibinfo{year}{2017}\natexlab{}.
\newblock \showarticletitle{A unified approach to interpreting model
  predictions}. In \bibinfo{booktitle}{\emph{Proc.~of NIPS}}.
  \bibinfo{pages}{4768--4777}.
\newblock


\bibitem[Luong et~al\mbox{.}(2019)]%
        {comst-drl-networking}
\bibfield{author}{\bibinfo{person}{Nguyen~Cong Luong},
  \bibinfo{person}{Dinh~Thai Hoang}, \bibinfo{person}{Shimin Gong},
  \bibinfo{person}{Dusit Niyato}, \bibinfo{person}{Ping Wang},
  \bibinfo{person}{Ying-Chang Liang}, {and} \bibinfo{person}{Dong~In Kim}.}
  \bibinfo{year}{2019}\natexlab{}.
\newblock \showarticletitle{Applications of Deep Reinforcement Learning in
  Communications and Networking: A Survey}.
\newblock \bibinfo{journal}{\emph{IEEE Communications Surveys \& Tutorials}}
  \bibinfo{volume}{21}, \bibinfo{number}{4} (\bibinfo{year}{2019}),
  \bibinfo{pages}{3133--3174}.
\newblock
\urldef\tempurl%
\url{https://doi.org/10.1109/COMST.2019.2916583}
\showDOI{\tempurl}


\bibitem[Madumal et~al\mbox{.}(2020)]%
        {madumal-xlr-casual-lens}
\bibfield{author}{\bibinfo{person}{Prashan Madumal}, \bibinfo{person}{Tim
  Miller}, \bibinfo{person}{Liz Sonenberg}, {and} \bibinfo{person}{Frank
  Vetere}.} \bibinfo{year}{2020}\natexlab{}.
\newblock \showarticletitle{Explainable Reinforcement Learning through a Causal
  Lens}.
\newblock \bibinfo{journal}{\emph{Proc.~of~AAAI Conference on Artificial
  Intelligence}} \bibinfo{volume}{34}, \bibinfo{number}{03}
  (\bibinfo{date}{Apr.} \bibinfo{year}{2020}), \bibinfo{pages}{2493--2500}.
\newblock
\urldef\tempurl%
\url{https://doi.org/10.1609/aaai.v34i03.5631}
\showDOI{\tempurl}


\bibitem[Mahimkar et~al\mbox{.}(2021)]%
        {auric}
\bibfield{author}{\bibinfo{person}{Ajay Mahimkar}, \bibinfo{person}{Ashiwan
  Sivakumar}, \bibinfo{person}{Zihui Ge}, \bibinfo{person}{Shomik Pathak},
  {and} \bibinfo{person}{Karunasish Biswas}.} \bibinfo{year}{2021}\natexlab{}.
\newblock \showarticletitle{Auric: Using Data-Driven Recommendation to
  Automatically Generate Cellular Configuration}. In
  \bibinfo{booktitle}{\emph{Proc.~of the ACM SIGCOMM}}.
  \bibinfo{pages}{807–820}.
\newblock
\showISBNx{9781450383837}
\urldef\tempurl%
\url{https://doi.org/10.1145/3452296.3472906}
\showDOI{\tempurl}


\bibitem[Mao et~al\mbox{.}(2017)]%
        {mao-pensieve-sigcomm17}
\bibfield{author}{\bibinfo{person}{Hongzi Mao}, \bibinfo{person}{Ravi
  Netravali}, {and} \bibinfo{person}{Mohammad Alizadeh}.}
  \bibinfo{year}{2017}\natexlab{}.
\newblock \showarticletitle{Neural Adaptive Video Streaming with Pensieve}. In
  \bibinfo{booktitle}{\emph{Proc.~of~ACM SIGCOMM}}. \bibinfo{pages}{197–210}.
\newblock
\showISBNx{9781450346535}
\urldef\tempurl%
\url{https://doi.org/10.1145/3098822.3098843}
\showDOI{\tempurl}


\bibitem[{Mavenir}(2023)]%
        {mavenir-drl}
\bibfield{author}{\bibinfo{person}{{Mavenir}}.}
  \bibinfo{year}{2023}\natexlab{}.
\newblock \bibinfo{title}{{Building the World's First O-RAN-Compliant,
  AI-Powered, Closed-Loop Near-RT RIC}}.
\newblock
  \bibinfo{howpublished}{\url{https://www.mavenir.com/blog/building-the-worlds-first-o-ran-compliant-ai-powered-closed-loop-near-rt-ric/}}.
\newblock


\bibitem[Melodia et~al\mbox{.}(2021)]%
        {colosseum}
\bibfield{author}{\bibinfo{person}{Tommaso Melodia}, \bibinfo{person}{Stefano
  Basagni}, \bibinfo{person}{Kaushik~R. Chowdhury}, \bibinfo{person}{Abhimanyu
  Gosain}, \bibinfo{person}{Michele Polese}, \bibinfo{person}{Pedram Johari},
  {and} \bibinfo{person}{Leonardo Bonati}.} \bibinfo{year}{2021}\natexlab{}.
\newblock \showarticletitle{Colosseum, the World's Largest Wireless Network
  Emulator}. In \bibinfo{booktitle}{\emph{Proc.~of~ACM MobiCom}}.
  \bibinfo{pages}{860–861}.
\newblock
\showISBNx{9781450383424}
\urldef\tempurl%
\url{https://doi.org/10.1145/3447993.3488032}
\showDOI{\tempurl}


\bibitem[Meng et~al\mbox{.}(2020)]%
        {zili-interpret-sigcomm20}
\bibfield{author}{\bibinfo{person}{Zili Meng}, \bibinfo{person}{Minhu Wang},
  \bibinfo{person}{Jiasong Bai}, \bibinfo{person}{Mingwei Xu},
  \bibinfo{person}{Hongzi Mao}, {and} \bibinfo{person}{Hongxin Hu}.}
  \bibinfo{year}{2020}\natexlab{}.
\newblock \showarticletitle{Interpreting Deep Learning-Based Networking
  Systems}. In \bibinfo{booktitle}{\emph{Proc.~of~ACM~SIGCOMM}}.
  \bibinfo{pages}{154–171}.
\newblock
\showISBNx{9781450379557}
\urldef\tempurl%
\url{https://doi.org/10.1145/3387514.3405859}
\showDOI{\tempurl}


\bibitem[Middleton et~al\mbox{.}(2022)]%
        {regulating-ai}
\bibfield{author}{\bibinfo{person}{Stuart~E. Middleton},
  \bibinfo{person}{Emmanuel Letouz\'{e}}, \bibinfo{person}{Ali Hossaini}, {and}
  \bibinfo{person}{Adriane Chapman}.} \bibinfo{year}{2022}\natexlab{}.
\newblock \showarticletitle{Trust, Regulation, and Human-in-the-Loop {AI}:
  Within the European Region}.
\newblock \bibinfo{journal}{\emph{Commun. ACM}} \bibinfo{volume}{65},
  \bibinfo{number}{4} (\bibinfo{date}{mar} \bibinfo{year}{2022}),
  \bibinfo{pages}{64–68}.
\newblock
\showISSN{0001-0782}


\bibitem[Montavon et~al\mbox{.}(2019)]%
        {montavon-lrp}
\bibfield{author}{\bibinfo{person}{Gr{\'e}goire Montavon},
  \bibinfo{person}{Alexander Binder}, \bibinfo{person}{Sebastian Lapuschkin},
  \bibinfo{person}{Wojciech Samek}, {and} \bibinfo{person}{Klaus-Robert
  M{\"u}ller}.} \bibinfo{year}{2019}\natexlab{}.
\newblock \bibinfo{booktitle}{\emph{Layer-Wise Relevance Propagation: An
  Overview}}.
\newblock \bibinfo{publisher}{Springer International Publishing},
  \bibinfo{pages}{193--209}.
\newblock
\urldef\tempurl%
\url{https://doi.org/10.1007/978-3-030-28954-6\_10}
\showDOI{\tempurl}


\bibitem[Moulay et~al\mbox{.}(2022)]%
        {moulay2022automated}
\bibfield{author}{\bibinfo{person}{Mohamed Moulay},
  \bibinfo{person}{Rafael~Garcia Leiva}, \bibinfo{person}{Pablo J~Rojo Maroni},
  \bibinfo{person}{Fernando Diez}, \bibinfo{person}{Vincenzo Mancuso}, {and}
  \bibinfo{person}{Antonio~Fernandez Anta}.} \bibinfo{year}{2022}\natexlab{}.
\newblock \showarticletitle{Automated identification of network anomalies and
  their causes with interpretable machine learning: The {CIAN} methodology and
  {TTrees} implementation}.
\newblock \bibinfo{journal}{\emph{Computer Communications}}
  \bibinfo{volume}{191} (\bibinfo{year}{2022}), \bibinfo{pages}{327--348}.
\newblock


\bibitem[Naseer and Benson(2022)]%
        {naseer2022configanator}
\bibfield{author}{\bibinfo{person}{Usama Naseer} {and}
  \bibinfo{person}{Theophilus~A Benson}.} \bibinfo{year}{2022}\natexlab{}.
\newblock \showarticletitle{Configanator: A Data-driven Approach to Improving
  {CDN} Performance}. In \bibinfo{booktitle}{\emph{Proc.~of~USENIX NSDI}}.
  \bibinfo{pages}{1135--1158}.
\newblock
\showISBNx{978-1-939133-27-4}


\bibitem[{O-RAN Software Community}({[n.\,d.]})]%
        {amber_release}
\bibfield{author}{\bibinfo{person}{{O-RAN Software Community}}.}
  \bibinfo{year}{[n.\,d.]}\natexlab{}.
\newblock \bibinfo{title}{Amber Release}.
\newblock
  \bibinfo{howpublished}{\url{https://wiki.o-ran-sc.org/pages/viewpage.action?pageId=14221337}}.
\newblock
\newblock
\shownote{Accessed July 2020}.


\bibitem[{O-RAN Working Group 3}(2021)]%
        {oran-wg3-ricarch}
\bibfield{author}{\bibinfo{person}{{O-RAN Working Group 3}}.}
  \bibinfo{year}{2021}\natexlab{}.
\newblock \bibinfo{title}{{O-RAN Near-RT RAN Intelligent Controller Near-RT RIC
  Architecture 2.00}}.
\newblock \bibinfo{howpublished}{O-RAN.WG3.RICARCH-v02.00}.
\newblock


\bibitem[Polese et~al\mbox{.}(2022)]%
        {tmc-coloran-wines}
\bibfield{author}{\bibinfo{person}{Michele Polese}, \bibinfo{person}{Leonardo
  Bonati}, \bibinfo{person}{Salvatore D'Oro}, \bibinfo{person}{Stefano
  Basagni}, {and} \bibinfo{person}{Tommaso Melodia}.}
  \bibinfo{year}{2022}\natexlab{}.
\newblock \showarticletitle{{ColO-RAN}: Developing Machine Learning-based
  {xApps} for {Open RAN} Closed-loop Control on Programmable Experimental
  Platforms}.
\newblock \bibinfo{journal}{\emph{IEEE Transactions on Mobile Computing}}
  (\bibinfo{year}{2022}), \bibinfo{pages}{1--14}.
\newblock
\urldef\tempurl%
\url{https://doi.org/10.1109/TMC.2022.3188013}
\showDOI{\tempurl}


\bibitem[Polese et~al\mbox{.}(2023)]%
        {comst-surveoran-wines}
\bibfield{author}{\bibinfo{person}{Michele Polese}, \bibinfo{person}{Leonardo
  Bonati}, \bibinfo{person}{Salvatore D’Oro}, \bibinfo{person}{Stefano
  Basagni}, {and} \bibinfo{person}{Tommaso Melodia}.}
  \bibinfo{year}{2023}\natexlab{}.
\newblock \showarticletitle{Understanding O-RAN: Architecture, Interfaces,
  Algorithms, Security, and Research Challenges}.
\newblock \bibinfo{journal}{\emph{IEEE Communications Surveys \& Tutorials}}
  (\bibinfo{year}{2023}), \bibinfo{pages}{1--1}.
\newblock
\urldef\tempurl%
\url{https://doi.org/10.1109/COMST.2023.3239220}
\showDOI{\tempurl}


\bibitem[Prakash et~al\mbox{.}(2019)]%
        {ac-drl-example}
\bibfield{author}{\bibinfo{person}{Bharat Prakash}, \bibinfo{person}{Mark
  Horton}, \bibinfo{person}{Nicholas~R. Waytowich},
  \bibinfo{person}{William~David Hairston}, \bibinfo{person}{Tim Oates}, {and}
  \bibinfo{person}{Tinoosh Mohsenin}.} \bibinfo{year}{2019}\natexlab{}.
\newblock \showarticletitle{On the Use of Deep Autoencoders for Efficient
  Embedded Reinforcement Learning}. In \bibinfo{booktitle}{\emph{Proc.~of ACM
  GLSVLSI}}. \bibinfo{pages}{507–512}.
\newblock
\showISBNx{9781450362528}
\urldef\tempurl%
\url{https://doi.org/10.1145/3299874.3319493}
\showDOI{\tempurl}


\bibitem[Puiutta and Veith(2020)]%
        {puiatta-xrl-survey}
\bibfield{author}{\bibinfo{person}{Erika Puiutta} {and} \bibinfo{person}{Eric
  M. S.~P. Veith}.} \bibinfo{year}{2020}\natexlab{}.
\newblock \showarticletitle{Explainable Reinforcement Learning: A Survey}. In
  \bibinfo{booktitle}{\emph{Machine Learning and Knowledge Extraction}},
  \bibfield{editor}{\bibinfo{person}{Andreas Holzinger}, \bibinfo{person}{Peter
  Kieseberg}, \bibinfo{person}{A~Min Tjoa}, {and} \bibinfo{person}{Edgar
  Weippl}} (Eds.). \bibinfo{publisher}{Springer International Publishing},
  \bibinfo{pages}{77--95}.
\newblock
\showISBNx{978-3-030-57321-8}


\bibitem[Ribeiro et~al\mbox{.}(2016)]%
        {lime-kdd}
\bibfield{author}{\bibinfo{person}{Marco~Tulio Ribeiro},
  \bibinfo{person}{Sameer Singh}, {and} \bibinfo{person}{Carlos Guestrin}.}
  \bibinfo{year}{2016}\natexlab{}.
\newblock \showarticletitle{{``Why Should I Trust You?''}: Explaining the
  Predictions of Any Classifier}. In \bibinfo{booktitle}{\emph{Proc.~of ACM
  SIGKDD}}. \bibinfo{pages}{1135–1144}.
\newblock
\showISBNx{9781450342322}
\urldef\tempurl%
\url{https://doi.org/10.1145/2939672.2939778}
\showDOI{\tempurl}


\bibitem[Scapin et~al\mbox{.}(2022)]%
        {shap-biomedic}
\bibfield{author}{\bibinfo{person}{Daniele Scapin}, \bibinfo{person}{Giulia
  Cisotto}, \bibinfo{person}{Elvina Gindullina}, {and}
  \bibinfo{person}{Leonardo Badia}.} \bibinfo{year}{2022}\natexlab{}.
\newblock \showarticletitle{Shapley Value as an Aid to Biomedical Machine
  Learning: a Heart Disease Dataset Analysis}. In
  \bibinfo{booktitle}{\emph{Proc.~of~IEEE CCGrid}}. \bibinfo{pages}{933--939}.
\newblock
\urldef\tempurl%
\url{https://doi.org/10.1109/CCGrid54584.2022.00113}
\showDOI{\tempurl}


\bibitem[Serly et~al\mbox{.}(2023)]%
        {deexp23}
\bibfield{author}{\bibinfo{person}{Moghadas~Gholian Serly},
  \bibinfo{person}{Fiandrino Claudio}, \bibinfo{person}{Collet Alan},
  \bibinfo{person}{Attanasio Giulia}, \bibinfo{person}{Fiore Marco}, {and}
  \bibinfo{person}{Widmer Joerg}.} \bibinfo{year}{2023}\natexlab{}.
\newblock \showarticletitle{Spotting Deep Neural Network Vulnerabilities in
  Mobile Traffic Forecasting with an Explainable {AI} Lens}. In
  \bibinfo{booktitle}{\emph{Proc.~of IEEE INFOCOM}}. \bibinfo{pages}{1--10}.
\newblock
\urldef\tempurl%
\url{https://doi.org/10.1109/INFOCOM53939.2023.10228989}
\showDOI{\tempurl}


\bibitem[Shrikumar et~al\mbox{.}(2017)]%
        {deeplift}
\bibfield{author}{\bibinfo{person}{Avanti Shrikumar}, \bibinfo{person}{Peyton
  Greenside}, {and} \bibinfo{person}{Anshul Kundaje}.}
  \bibinfo{year}{2017}\natexlab{}.
\newblock \showarticletitle{Learning important features through propagating
  activation differences}. In \bibinfo{booktitle}{\emph{Proc.~of~ICML}}.
  \bibinfo{pages}{3145--3153}.
\newblock


\bibitem[Smith et~al\mbox{.}(2021)]%
        {milcom21-ss}
\bibfield{author}{\bibinfo{person}{Rob Smith}, \bibinfo{person}{Connor
  Freeberg}, \bibinfo{person}{Travis Machacek}, {and}
  \bibinfo{person}{Venkatesh Ramaswamy}.} \bibinfo{year}{2021}\natexlab{}.
\newblock \showarticletitle{An {O-RAN} Approach to Spectrum Sharing Between
  Commercial {5G} and Government Satellite Systems}. In
  \bibinfo{booktitle}{\emph{Proc.~of~IEEE MILCOM}}. \bibinfo{pages}{739--744}.
\newblock
\urldef\tempurl%
\url{https://doi.org/10.1109/MILCOM52596.2021.9653140}
\showDOI{\tempurl}


\bibitem[Sutton and Barto(2018)]%
        {sutton2018reinforcement}
\bibfield{author}{\bibinfo{person}{Richard~S Sutton} {and}
  \bibinfo{person}{Andrew~G Barto}.} \bibinfo{year}{2018}\natexlab{}.
\newblock \showarticletitle{Reinforcement learning: An introduction second
  edition}.
\newblock \bibinfo{journal}{\emph{Adaptive computation and machine learning:
  The MIT Press, Cambridge, MA and London}} (\bibinfo{year}{2018}).
\newblock


\bibitem[{Unwired Labs}(2023)]%
        {opencellid}
\bibfield{author}{\bibinfo{person}{{Unwired Labs}}.} \bibinfo{year}{Accessed
  March 2023}\natexlab{}.
\newblock \bibinfo{title}{{OpenCelliD}}.
\newblock \bibinfo{howpublished}{\url{https://opencellid.org}}.
\newblock


\bibitem[{U.S. Naval Research Laboratory}({[n.\,d.]})]%
        {mgen}
\bibfield{author}{\bibinfo{person}{{U.S. Naval Research Laboratory}}.}
  \bibinfo{year}{[n.\,d.]}\natexlab{}.
\newblock \bibinfo{title}{{MGEN Traffic Emulator}}.
\newblock
  \bibinfo{howpublished}{\url{https://www.nrl.navy.mil/Our-Work/Areas-of-Research/Information-Technology/NCS/MGEN}}.
\newblock
\newblock
\shownote{Accessed September 2021}.


\bibitem[Wang et~al\mbox{.}(2021)]%
        {arxiv-survey}
\bibfield{author}{\bibinfo{person}{Shen Wang}, \bibinfo{person}{M.~Atif
  Qureshi}, \bibinfo{person}{Luis Miralles-Pechuán}, \bibinfo{person}{Thien
  Huynh-The}, \bibinfo{person}{Thippa~Reddy Gadekallu}, {and}
  \bibinfo{person}{Madhusanka Liyanage}.} \bibinfo{year}{2021}\natexlab{}.
\newblock \bibinfo{title}{Explainable {AI} for {B5G/6G}: Technical Aspects, Use
  Cases, and Research Challenges}.
\newblock
\newblock
\urldef\tempurl%
\url{https://doi.org/10.48550/ARXIV.2112.04698}
\showDOI{\tempurl}


\bibitem[Xu et~al\mbox{.}(2022)]%
        {xu-lrp-drl}
\bibfield{author}{\bibinfo{person}{Rui Xu}, \bibinfo{person}{Siyu Luan},
  \bibinfo{person}{Zonghua Gu}, \bibinfo{person}{Qingling Zhao}, {and}
  \bibinfo{person}{Gang Chen}.} \bibinfo{year}{2022}\natexlab{}.
\newblock \showarticletitle{{LRP}-based Policy Pruning and Distillation of
  Reinforcement Learning Agents for Embedded Systems}. In
  \bibinfo{booktitle}{\emph{Proc.~of~IEEE ISORC}}. \bibinfo{pages}{1--8}.
\newblock
\urldef\tempurl%
\url{https://doi.org/10.1109/ISORC52572.2022.9812837}
\showDOI{\tempurl}


\bibitem[Xu et~al\mbox{.}(2018)]%
        {xu-drl-infocom18}
\bibfield{author}{\bibinfo{person}{Zhiyuan Xu}, \bibinfo{person}{Jian Tang},
  \bibinfo{person}{Jingsong Meng}, \bibinfo{person}{Weiyi Zhang},
  \bibinfo{person}{Yanzhi Wang}, \bibinfo{person}{Chi~Harold Liu}, {and}
  \bibinfo{person}{Dejun Yang}.} \bibinfo{year}{2018}\natexlab{}.
\newblock \showarticletitle{Experience-driven Networking: A Deep Reinforcement
  Learning based Approach}. In \bibinfo{booktitle}{\emph{Proc.~of IEEE
  INFOCOM}}. \bibinfo{pages}{1871--1879}.
\newblock
\urldef\tempurl%
\url{https://doi.org/10.1109/INFOCOM.2018.8485853}
\showDOI{\tempurl}


\bibitem[Yu et~al\mbox{.}(2022)]%
        {tarik-drl-routing-22}
\bibfield{author}{\bibinfo{person}{Hao Yu}, \bibinfo{person}{Tarik Taleb},
  {and} \bibinfo{person}{Jiawei Zhang}.} \bibinfo{year}{2022}\natexlab{}.
\newblock \showarticletitle{Deep Reinforcement Learning based Deterministic
  Routing and Scheduling for Mixed-Criticality Flows}.
\newblock \bibinfo{journal}{\emph{IEEE Transactions on Industrial Informatics}}
  (\bibinfo{year}{2022}), \bibinfo{pages}{1--11}.
\newblock
\urldef\tempurl%
\url{https://doi.org/10.1109/TII.2022.3222314}
\showDOI{\tempurl}


\bibitem[Zhang et~al\mbox{.}(2017)]%
        {zhang-hypergraphs-5g}
\bibfield{author}{\bibinfo{person}{Hongliang Zhang}, \bibinfo{person}{Lingyang
  Song}, \bibinfo{person}{Yonghui Li}, {and} \bibinfo{person}{Geoffrey~Ye Li}.}
  \bibinfo{year}{2017}\natexlab{}.
\newblock \showarticletitle{Hypergraph Theory: Applications in {5G}
  Heterogeneous Ultra-Dense Networks}.
\newblock \bibinfo{journal}{\emph{IEEE Communications Magazine}}
  \bibinfo{volume}{55}, \bibinfo{number}{12} (\bibinfo{year}{2017}),
  \bibinfo{pages}{70--76}.
\newblock
\urldef\tempurl%
\url{https://doi.org/10.1109/MCOM.2017.1700400}
\showDOI{\tempurl}


\bibitem[Zhang et~al\mbox{.}(2022)]%
        {zhang-shap-drl-power-tcss}
\bibfield{author}{\bibinfo{person}{Ke Zhang}, \bibinfo{person}{Jun Zhang},
  \bibinfo{person}{Pei-Dong Xu}, \bibinfo{person}{Tianlu Gao}, {and}
  \bibinfo{person}{David~Wenzhong Gao}.} \bibinfo{year}{2022}\natexlab{}.
\newblock \showarticletitle{{Explainable AI} in Deep Reinforcement Learning
  Models for Power System Emergency Control}.
\newblock \bibinfo{journal}{\emph{IEEE Transactions on Computational Social
  Systems}} \bibinfo{volume}{9}, \bibinfo{number}{2} (\bibinfo{year}{2022}),
  \bibinfo{pages}{419--427}.
\newblock
\urldef\tempurl%
\url{https://doi.org/10.1109/TCSS.2021.3096824}
\showDOI{\tempurl}


\bibitem[Zheng and Li(2022)]%
        {infocom22zheng}
\bibfield{author}{\bibinfo{person}{Tianhang Zheng} {and}
  \bibinfo{person}{Baochun Li}.} \bibinfo{year}{2022}\natexlab{}.
\newblock \showarticletitle{Poisoning Attacks on Deep Learning based Wireless
  Traffic Prediction}. In \bibinfo{booktitle}{\emph{Proc.~of IEEE INFOCOM}}.
  \bibinfo{pages}{660–669}.
\newblock
\urldef\tempurl%
\url{https://doi.org/10.1109/INFOCOM48880.2022.9796791}
\showDOI{\tempurl}


\bibitem[Zheng et~al\mbox{.}(2022)]%
        {zheng-drl-jsac}
\bibfield{author}{\bibinfo{person}{Ying Zheng}, \bibinfo{person}{Lixiang Lin},
  \bibinfo{person}{Tianqi Zhang}, \bibinfo{person}{Haoyu Chen},
  \bibinfo{person}{Qingyang Duan}, \bibinfo{person}{Yuedong Xu}, {and}
  \bibinfo{person}{Xin Wang}.} \bibinfo{year}{2022}\natexlab{}.
\newblock \showarticletitle{Enabling Robust {DRL}-Driven Networking Systems via
  Teacher-Student Learning}.
\newblock \bibinfo{journal}{\emph{IEEE Journal on Selected Areas in
  Communications}} \bibinfo{volume}{40}, \bibinfo{number}{1}
  (\bibinfo{year}{2022}), \bibinfo{pages}{376--392}.
\newblock
\urldef\tempurl%
\url{https://doi.org/10.1109/JSAC.2021.3126085}
\showDOI{\tempurl}


\bibitem[Zheng et~al\mbox{.}(2018)]%
        {zheng2018demystifying}
\bibfield{author}{\bibinfo{person}{Ying Zheng}, \bibinfo{person}{Ziyu Liu},
  \bibinfo{person}{Xinyu You}, \bibinfo{person}{Yuedong Xu}, {and}
  \bibinfo{person}{Junchen Jiang}.} \bibinfo{year}{2018}\natexlab{}.
\newblock \showarticletitle{Demystifying deep learning in networking}. In
  \bibinfo{booktitle}{\emph{Proc.~of~APNet}}. \bibinfo{pages}{1--7}.
\newblock


\end{thebibliography}

\clearpage
\appendix

\section{Experimental Configurations}
\label{app-sec:exp-config}

\begin{table*}[tbh]
\centering
\caption{The configurations utilized in the experiments\vspace{-.15cm}}%
\label{tab:exp-settings}%
\vspace*{-2ex}%
\resizebox{\textwidth}{!}{%
\begin{tabular}{lc|cccccc|cccc}
	\toprule
	\multirow{2}{*}{\textsc{Agent} } & \textsc{Traf.} & \multicolumn{6}{c|}{\textsc{Num. Users}} & \multicolumn{4}{c}{\textsc{Action Steering Strategy} (Users: 6, drop to 5)}
	\\
	\cmidrule(lr){3-8}\cmidrule(lr){9-12}
	& \textsc{Scen.} 
	& 6 & 5 & 4 & 3 & 2 & 1 &
	\ref{ar:1}  & \ref{ar:2} & \ref{ar:3} & \textsc{Baseline} \\
	\midrule
	\multirow{2}{*}{\ac{ht}} & TRF1 
	& C$_{\text{\ac{ht},trf1}-6}$ & C$_{\text{\ac{ht},trf1}-5}$ & C$_{\text{\ac{ht},trf1}-4}$ & C$_{\text{\ac{ht},trf1}-3}$ & C$_{\text{\ac{ht},trf1}-2}$ & C$_{\text{\ac{ht},trf1}-1}$ 
	& C$_{\text{\ac{ht},trf1-a1-10}}$, C$_{\text{\ac{ht},trf1-a1-20}}$ & C$_{\text{\ac{ht},trf1-a2-10}}$, C$_{\text{\ac{ht},trf1-a2-20}}$ & C$_{\text{\ac{ht},trf1-a3-10}}$, C$_{\text{\ac{ht},trf1-a3-20}}$ & C$_{\text{\ac{ht},trf1-b-10}}$, C$_{\text{\ac{ht},trf1-b-20}}$ \\
	& TRF2 & C$_{\text{\ac{ht},trf2}-6}$ & C$_{\text{\ac{ht},trf2}-5}$ & C$_{\text{\ac{ht},trf2}-4}$ & C$_{\text{\ac{ht},trf2}-3}$ & C$_{\text{\ac{ht},trf2}-2}$ & C$_{\text{\ac{ht},trf2}-1}$ 
	& C$_{\text{\ac{ht},trf2-a1-10}}$, C$_{\text{\ac{ht},trf2-a1-20}}$ & C$_{\text{\ac{ht},trf2-a2-10}}$, C$_{\text{\ac{ht},trf2-a2-20}}$ & C$_{\text{\ac{ht},trf2-a3-10}}$, C$_{\text{\ac{ht},trf2-a3-20}}$ & C$_{\text{\ac{ht},trf2-b-10}}$, C$_{\text{\ac{ht},trf2-b-20}}$\\
	\multirow{2}{*}{\ac{ll}} & TRF1 & C$_{\text{\ac{ll},trf1}-6}$ & C$_{\text{\ac{ll},trf1}-5}$ & C$_{\text{\ac{ll},trf1}-4}$ & C$_{\text{\ac{ll},trf1}-3}$ & C$_{\text{\ac{ll},trf1}-2}$ & C$_{\text{\ac{ll},trf1}-1}$ 
	& C$_{\text{\ac{ll},trf1-a1-10}}$, C$_{\text{\ac{ll},trf1-a1-20}}$ & C$_{\text{\ac{ll},trf1-a2-10}}$, C$_{\text{\ac{ll},trf1-a2-20}}$ & C$_{\text{\ac{ll},trf1-a3-10}}$, C$_{\text{\ac{ll},trf1-a3-20}}$ & C$_{\text{\ac{ll},trf1-b-10}}$, C$_{\text{\ac{ll},trf1-b-20}}$ \\
	& TRF2 & C$_{\text{\ac{ll},trf2}-6}$ & C$_{\text{\ac{ll},trf2}-5}$ & C$_{\text{\ac{ll},trf2}-4}$ & C$_{\text{\ac{ll},trf2}-3}$ & C$_{\text{\ac{ll},trf2}-2}$ & C$_{\text{\ac{ll},trf2}-1}$ 
	& C$_{\text{\ac{ll},trf2-a1-10}}$, C$_{\text{\ac{ll},trf2-a1-20}}$ & C$_{\text{\ac{ll},trf2-a2-10}}$, C$_{\text{\ac{ll},trf2-a2-20}}$ & C$_{\text{\ac{ll},trf2-a3-10}}$, C$_{\text{\ac{ll},trf2-a3-20}}$ & C$_{\text{\ac{ll},trf2-b-10}}$, C$_{\text{\ac{ll},trf2-b-20}}$ \\
	\bottomrule
\end{tabular}
}
\vspace*{-2ex}%
\end{table*}

Table~\ref{tab:exp-settings} shows the configurations C used in the experiments (\S~\ref{subsec:motivation-shap} and \S~\ref{sec:experiments}) for different agents and different traffic scenarios. Overall, we run 48 different experiments. Each configuration in the left part of the table refers to experiments with different numbers of users, \eg the tuple ``HT,tr1-6'' indicates a configuration C with the \ac{ht} agent, traffic profile 1, and 6 users. Users are equally assigned to each slice in the experiments with 6 and 3 users. For the experiments with 5, 4 and 2 users we use the following assignment:
\begin{itemize}
\item 5 users: 2 users to the slice \ac{embb}, 1 user to the slice \ac{mmtc} and 2 users to the slice \ac{urllc};
\item 4 users: 1 user to the slice \ac{embb}, 1 user to the slice \ac{mmtc} and 2 users to the slice \ac{urllc};
\item 2 users: 1 user to the slice \ac{embb}, no users to the slice \ac{mmtc} and 1 user to the slice \ac{urllc}.
\end{itemize}
The experiments with 1 user are executed three times, one for each of the the three slices \ac{embb}, \ac{mmtc}, \ac{urllc}. The right part of the table refe5rs to the experiments with intent-based action steering strategies (discussed in \S~\ref{subsec:results-optimization}), \eg the tuple ``\ac{ht},trf1-a1-10'' denotes a configuration C with the \ac{ht} agent, traffic profile 1, policy replacement \ref{ar:1}, and an observation window $O$ of 10 entries.

\section{Generating Attributed Graphs}
\label{app-sec:example-graph}
With the help of an example, we now explain how \toolname builds the attributed graph for the case of the agent \ac{ht} and traffic scenario TRF1. We first provide a step-by-step explanation and then show the overall graph.

\begin{figure*}
\centering%
\includegraphics[width=0.85\textwidth,keepaspectratio]{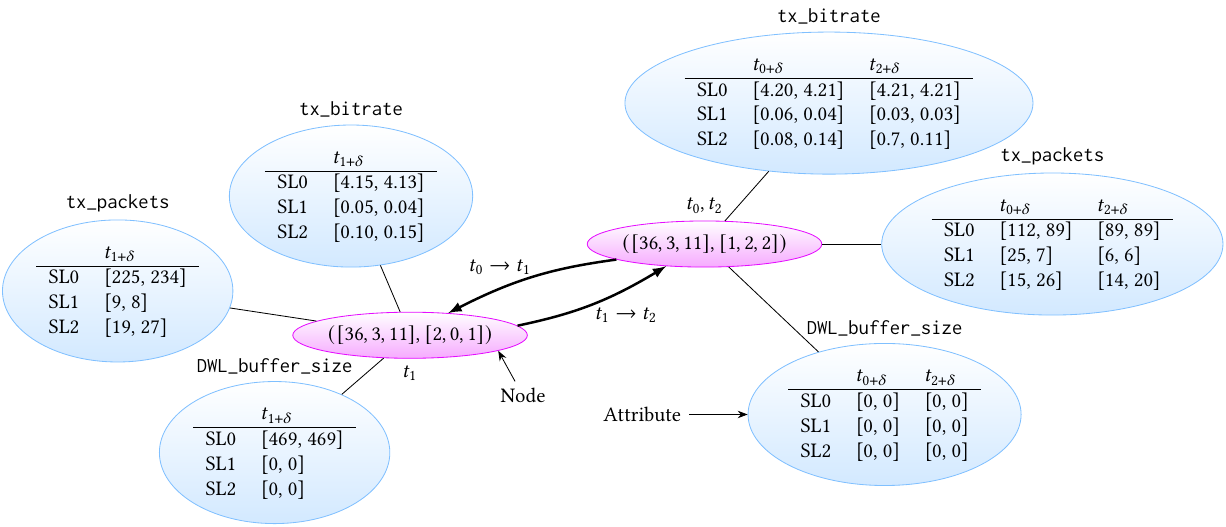}%
\vspace*{-3ex}%
\caption{Process of building the attributed graph during time steps $t_0$, $t_1$ and $t_2$}%
\label{fig:g-embb-trf1-3steps}%
\vspace*{-2ex}%
\end{figure*}

\begin{figure*}
\centering%
\includegraphics[width=0.85\textwidth,keepaspectratio]{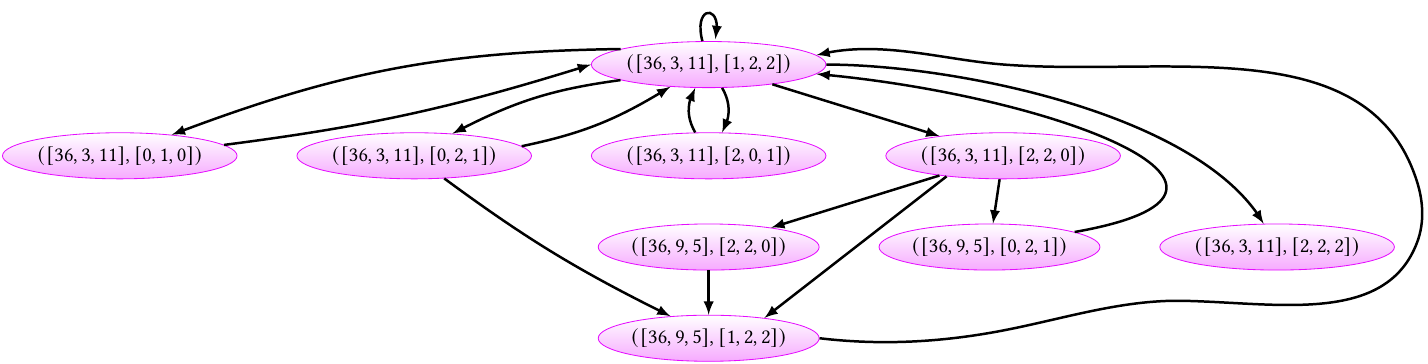}%
\vspace*{-2ex}%
\caption{The resulting graph for the \ac{ht} agent utilized in the presence of TRF1 traffic with a focus on actions}%
\label{fig:g-urllc-trf1}%
\vspace*{-2ex}%
\end{figure*}

In Figure~\ref{fig:g-embb-trf1-3steps}, we show both nodes (\ie the actions) and attributes during three consecutive time steps $t_0$, $t_1$, and $t_2$ in the observation window $W$. Two users per slice are present in the system for a total of six users. Each node in the graph contains a multi-modal action. The node \mbox{([36, 3, 11], [2, 0, 1])} represents the \ac{ran} slicing \ac{prb} allocation on the left, \ie \mbox{[36, 3, 11]} and scheduling policy allocation on the right, \ie \mbox{[2, 0, 1]} (0 denotes Round Robin (RR), 1 - Waterfilling (WF), and 2 - Proportional Fair (PF)). The positions inside the array refer to the slices. For example, \mbox{[36, 3, 11]} denotes an allocation of 36 \ac{prb} to the slice \ac{embb}, 3 to the slice \ac{mmtc}, and 11 to the slice \ac{urllc}. Each attribute in the graph contains the distribution of \acp{kpi} per slice of each user. For example, \mbox{SL0 [225,234]} of the attribute \txpkts indicates that two users have transmitted 225 and 234 packets for the slice 0. At time $t_0$, the agent enforces the action \mbox{([36, 3, 11], [1, 2, 2])}. We record in three different attributes, one per \ac{kpi} (i.e., \txbrate, \txpkts, and \dlbuff), the corresponding data observation for each slice (i.e., SL0, SL1, and SL2) and for each user. These attributes describe the impact of the action taken at $t_0$ on the future state $t_{0+\delta}$ (with $\delta=250$~ms) and that is taken as input by the \ac{drl} framework (autoencoder with \ac{drl} agent) to take action \mbox{([36, 3, 11], [2, 0, 1])} at $t_1$. As this is a new action, \toolname adds a new node, monitors the future state $t_{1+\delta}$ and records it in form of attributes. In the next iteration at $t_2$, the agent takes again the action \mbox{([36, 3, 11], [1, 2, 2])}. This is not a new action, hence \toolname updates the existing attributes recorded at $t_{0+\delta}$ with the new attributed at $t_{2+\delta}$. Throughout $W$, the attributes build in form of a distribution that contains as many samples as the occurrences of the action. The analysis of the distributions for each \ac{kpi} and slice between transitions allows to reveal the rationale for which the agent changes action. For example, in the transition from $t_1\rightarrow t_2$, the \txbrate for SL0 increases. Finally, in Figure~\ref{fig:g-urllc-trf1}, we show the complete graph and we only represent nodes and not attributes for readability.

\section{Considerations on the Agent LL}
\label{app-sec:agentll}

\begin{figure*}
\captionsetup[subfigure]{captionskip=0pt}
\centering
\hspace*{3em}\includegraphics[scale=.7]{./results-images/legend}\par
\vspace*{-2.5ex}%
\subfloat[\scriptsize \txbrate and \dlbuff~\label{fig:urllc-trf2-txbrate-dlbuffer}]{%
	\includegraphics[width=.3\textwidth,keepaspectratio]{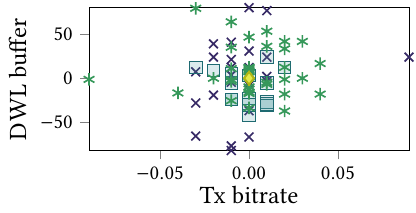}\vspace*{-2ex}%
}%
\quad%
\subfloat[\scriptsize \txpkts and \dlbuff~\label{fig:urllc-trf2-txpkts-dlbuffer}]{%
	\includegraphics[width=.3\textwidth,keepaspectratio]{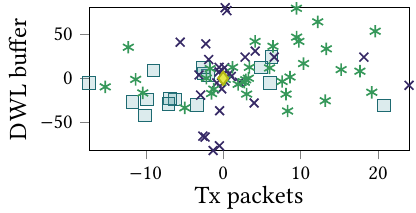}}%
\quad%
\subfloat[\scriptsize \txbrate and \txpkts~\label{fig:urllc-trf2-txbrate-txpkts}]{%
	\includegraphics[width=.3\textwidth,keepaspectratio]{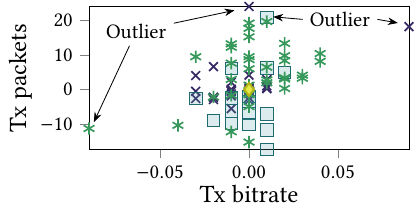}}%
\vspace*{-1.75ex}%
\caption{Detailed explanations for the \ac{ll} agent's behavior}%
\label{fig:explanation-urllc-trf2}%
\vspace*{-3ex}%
\end{figure*}

\begin{figure*}
\centering
\rotatebox{90}{%
	\includegraphics[width=0.95\textheight,keepaspectratio]{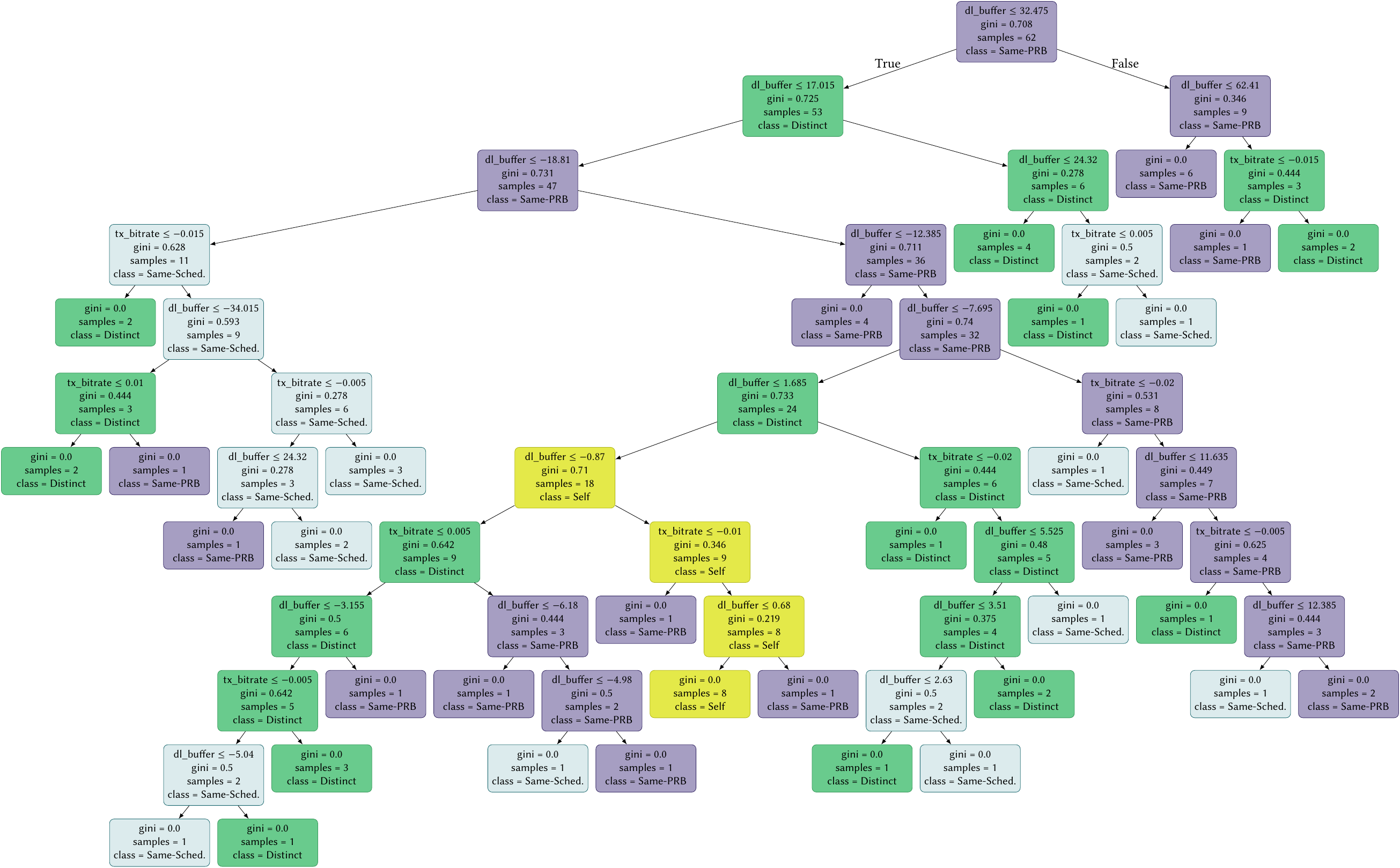}%
}%
\caption{\ac{dt} on \toolname explanations for the \ac{ll} agent}%
\label{fig:summary-of-exp-urllc-trf2}%
\end{figure*}

Figure~\ref{fig:explanation-urllc-trf2}, Figure~\ref{fig:summary-of-exp-urllc-trf2}, and Table~\ref{tab:summary-of-exp-urllc-trf2} show the process to obtain the summary of explanations for the agent \ac{ll} similarly to what Figure~\ref{fig:explanation-embb-trf1}, Figure~\ref{fig:summary-of-exp-embb-trf1}, and Table~\ref{tab:summary-of-exp-embb-trf1} represent for the agent \ac{ht}. With respect to the \ac{ht} counterpart, the agent \ac{ll} presents significant differences. 

\noindent$\bullet$ First, the number and magnitude of \ac{kpi} variations (for measurement units, refer to \S~\ref{subsec:drl-use-cases}) is, as expected, lower for the \ac{ll} agent than for the \ac{ht} agent. Indeed, the latter agent strives to guarantee high throughput for traffic flows with higher volume. The flows of the \ac{urllc} slice exhibit much lower volume. Maximizing throughput is comparatively easier than striving to minimize the buffer size (as proxy to low latency). This phenomenon can be observed in the higher number of transitions that the \ac{ll} agent performs if compared to its \ac{ht} counterpart.

\noindent$\bullet$ Second, note that the \ac{ll} agent shows a few outliers. These are transitions that generate an unexpected explanation. For example, in Figure~\ref{fig:explanation-urllc-trf2}\subref{fig:urllc-trf2-txbrate-txpkts} the ``Same-Sched.'' transition denoted as outlier is unexpected as all other transitions of such category usually lead to a decrease of \txpkts. We attribute this behavior to the fact that while traffic is deterministic in terms of statistical behavior, the actual instantaneous traffic load may deviate from the trend and thus produce outliers.

\noindent$\bullet$ Third, unlike the \ac{ht} agent that mainly uses the ``Same-PRB'' class ($40$\%), the \ac{ll} agent uses the two classes evenly.

\begin{table}
\centering
\caption{Summary of explanations for the \ac{ll} agent}%
\label{tab:summary-of-exp-urllc-trf2}%
\vspace*{-2ex}%
\resizebox{0.5\columnwidth}{!}{%
	\begin{tabular}{r>{\RaggedRight}p{7cm}}%
		\toprule
		\textsc{Transition} & \textsc{Interpretation}\\
		\midrule
		\textsc{Same-PRB} & Diminishes lightly \txbrate and augments \txpkts\\
		\textsc{Same-Sched.} & Diminishes \dlbuff, usually diminishes \txpkts and seldom marginally augments \txpkts \\
		\textsc{Distinct} & Mainly augments \txpkts \\
		\textsc{Self} & No change in \acp{kpi}\\
		\bottomrule
	\end{tabular}%
}%
\vspace*{-2ex}%
\end{table}

\section{Further Considerations on Action Steering}
\label{app-sec:action_steering}

\begin{figure}[h]
\captionsetup[subfigure]{captionskip=2pt}
\centering
\vspace*{-2.5ex}%
\subfloat[$O=10$~\label{fig:max-rew-obs-10}]{%
	\includegraphics[width=.45\textwidth,keepaspectratio]{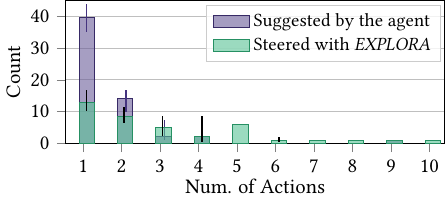}}%
~%
\subfloat[$O=20$~\label{fig:max-rew-obs-20}]{%
	\includegraphics[width=.45\textwidth,keepaspectratio]{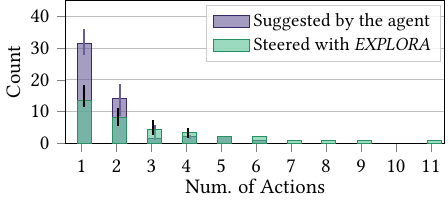}}%
\vspace*{-3ex}%
\caption{Distribution of actions with \ref{ar:1}}%
\label{fig:distr-action-replacement}%
\vspace*{-3ex}%
\end{figure}

We now assert that the intent-based action steering policies defined in \S~\ref{subsec:impl-ar-strategies} do not harm the capabilities of the agent to generalize and adapt to changes in the systems. Figure~\ref{fig:distr-action-replacement} portrays the median, first and third quartiles of the distributions obtained across all configuration settings for \ref{ar:1} (\ie \ac{ht} and \ac{ll} agents, TRF1 and TRF2 traffic scenarios) and varying the size of the past history $O$. We report both the number of times the attributed graph $G$ ``suggests'' to replace an action (purple bars), as well as the number of times such action is actually being replaced with the one suggested by the graph (green bars). 
Lower values of $O$ trigger comparatively a slightly higher number of action changes than higher values of $O$ (on average, $63$\% and $59$\%, respectively). However, through action changes, the agent probes less often new actions in a fully controlled fashion (the reduction between potentially used actions and those actually used is $25$\% for $O=10$ and $18$\% for $O=20$). The important remark from Figure~\ref{fig:distr-action-replacement} is that our intent-based action steering strategies are not preventing the agent to take a specific action (like shielding would do) because it is rare that the same action gets substituted more than 3 times. On the contrary, leveraging the explanations, the same actions suggested by the graph as replacement are used more often.

\section{Ethical Considerations}
This work does not raise any ethical issues as per the ACM Code of Ethics.

\end{document}